\documentclass[sigconf,screen=true]{acmart}

\renewcommand\footnotetextcopyrightpermission[1]{}
\usepackage{booktabs} % For formal tables

\usepackage{placeins} % For controlling the position of float images

\theoremstyle{plain}
\newtheorem{thrm}{Theorem}[section]

\theoremstyle{definition}
\newtheorem{prblm}[thrm]{Problem}

\usepackage[disable] %uncomment to disable messages
	{todonotes}
\newcommand{\todoVL}[1]{\todo[color=red!40, author=\textbf{Vincent}, inline]{#1}}

\newcommand{\todoNA}[1]{\todo[color=green!40, author=\textbf{Nejat}, inline]{#1}}
\hypersetup{draft}
\setcopyright{rightsretained}
\acmConference[i-Know 2017]{International Workshop on Social Network Analysis and Digital Humanities}{October 2017}{Graz, Austria} 
\acmYear{2017}
\copyrightyear{2017}

\begin{document}
\title{Signed Graph Analysis for the Interpretation of Voting Behavior}
%\titlenote{Produces the permission block, and
%  copyright information}
%\subtitle{Extended Abstract}
%\subtitlenote{The full version of the author's guide is available as
%  \texttt{acmart.pdf} document}

\author{Nejat Arinik}
\affiliation{%
  \institution{Laboratoire Informatique d'Avignon LIA EA 4128}
  \city{Avignon} 
  \country{France}
}
\email{nejat.arinik@univ-avignon.fr}

\author{Rosa Figueiredo}
\affiliation{%
  \institution{Laboratoire Informatique d'Avignon LIA EA 4128}
  \city{Avignon} 
  \country{France}
}
\email{rosa.figueiredo@univ-avignon.fr }

\author{Vincent Labatut}
\orcid{0000-0002-2619-2835}
\affiliation{%
  \institution{Laboratoire Informatique d'Avignon LIA EA 4128}
  \city{Avignon} 
  \country{France}
}
\email{vincent.labatut@univ-avignon.fr}


\begin{abstract}
In a \textit{signed graph}, each link is labeled with either a positive or a negative sign. This is particularly appropriate to model polarized systems. Such a graph can be characterized through the notion of \textit{structural balance}, which relies on the partitioning of the graph into internally solidary but mutually hostile subgroups. In this work, we show that signed graphs can be used to model and understand voting behavior. We take advantage of data from the European Parliament to confront two variants of structural balance, and illustrate how their use can help better understanding the studied system.
\end{abstract}




%% Used in some conference proceedings e.g. sigplan and sigchi
%   \caption{This is a teaser}

\maketitle

%%%%%%%%%%%%%%%%%%%%%%%%%%%%%%%%%%%%%%%%%%%%%%%%%%%%%%%%%%%%%%%
%%%%%%%%%%%%%%%%%%%%%%%%%%%%%%%%%%%%%%%%%%%%%%%%%%%%%%%%%%%%%%%
\section{Introduction}\label{intro}
\textit{Signed graphs} were primarily introduced in Psychology, with the objective of describing the relationships between people belonging to distinct social groups~\cite{Heider1946}: each one of their links is labeled with a sign $+$ or $-$, indicating the nature of the relationship between the considered adjacent nodes. A signed graph can be used to model any system containing two types of antithetical relationships, such as like/dislike, for/against, etc. Such a graph is considered \textit{structurally balanced} if it can be partitioned into two~\cite{Cartwright1956} or more~\cite{Davis1967} mutually hostile subgroups each having internal solidarity. Here, the words \textit{hostile} and \textit{solidarity} mean: \textit{connected by negative} and \textit{positive} links, respectively. 

However, it is very rare for a real-world network\footnote{We use the following words indistinctly in this article: \textit{graph} and \textit{network}, \textit{node} and \textit{vertex}, \textit{link} and \textit{edge}.} to have a perfectly balanced structure: the question is then to quantify how imbalanced it is. Various measures have been defined for this purpose, the simplest consisting in counting the numbers of misplaced links, i.e. positive ones located inside the groups, and negative ones located between them \cite{Cartwright1956}. Such measures are expressed relatively to a graph partition, so processing the graph balance amounts to identifying the partition corresponding to the lowest imbalance measure. In other words, calculating the graph balance can be formulated as an optimization problem. This type of optimization problem can be compared to that of \textit{community detection}, which consists in partitioning \textit{unsigned} networks in order to detect groups of nodes more densely connected relatively to the rest of the network \cite{Fortunato2010}. The main difference is of course the presence of signs attached to links, which represent additional information one has to take into account. Doing so is a non-trivial task, which cannot be conducted by simply performing minor adaptations of community detection methods \cite{Chiang2014}. %Moreover, even the notion of graph partitioning can be translated in several ways in the context of signed graphs, as we will see in Section~\ref{sec:PartioningMethodsSignedGraphs}.

In a previous article, we have studied the structural balance of weighted signed graphs representing the voting activity of Members of the European Parliament (MEPs) \cite{Mendonca2015}. We have compared the results obtained with partitioning methods designed for community detection on the one hand, and for structural balance on the other hand. Our main result was that, in contradiction with some conclusions presented in another study \cite{Esmailian2014}, taking negative links into account leads to significantly different partitions, at least for these data. This is consistent with other results appearing in the literature and showing that the information conveyed by negative links generally improves the resolution of the problems at hand, e.g. graph partitioning \cite{Doreian2015} or link prediction \cite{Kunegis2013}.

But our study suffers from several limitations. First, the extracted graphs contain many links with a close to zero weight, which could have been considered as noise: they largely increase the processing time and apparently make it harder to interpret the results, without significantly affecting the obtained partitions. Second, we focused our discussion on objective aspects (the quality of the partitions in terms of imbalance) and stayed quite superficial when interpreting our results relatively to the studied system. Third, the raw data we used as a base to extract the signed graphs were incomplete: they did not cover the whole 7th term.

In a more recent work also conducted in our research group \cite{Levorato2017}, Levorato \& Frota applied the same methodology as in \cite{Mendonca2015} and tried to solve certain of these limitations. They focused on a different dataset, representing the voting activity at the National Congress of Brazil, and performed a thorough interpretation of the obtained partitions. They also considered a variant of the structural balance, which we will describe later. However, they used the extraction approach from \cite{Mendonca2015}, leading to \textit{complete}, and therefore noisy, signed graphs. This may be the reason why certain of the partitions they obtained are marginally informative and/or difficult to interpret. Moreover, this difficulty is further increased by the current confused political situation in Brazil\footnote{\url{http://www.bbc.com/news/world-latin-america-35810578.}}.

In this paper, we present the work we conducted to overcome the limitations of both studies \cite{Mendonca2015, Levorato2017}, with the following contributions. First, we come back to the European Parliament (EP), through a different, complete, data source. Second, we include a filtering step to get rid of the noise present in the complete graphs. Third, we focus our interpretation only on a small part of the available data, in order to deliver a deeper analysis and better show the interest of structural balance to characterize and understand the considered system.

The rest of the article is organized as follows. In Section~\ref{sec:Methods}, we describe the methods we used to extract the signed networks from our raw data, as well as the partitioning algorithms we applied to them. In Section~\ref{sec:Stats}, we study the effect of the additional filtering step on these algorithms, and the quality of a heuristic proposed to measure the imbalance. In Section~\ref{sec:Interpretation}, we present our results on a few specific cases selected from our corpus, and discuss them. We adopt the perspective of the end-user, and instead of simply focusing on the objective performance of the considered partitioning algorithms, we also comment on the quality of the identified clusters relatively to the studied system. Finally, in Section~\ref{sec:Conclusion} we summarize our findings, comment the limitations of our work and describe how they can be overcome, and how our methods can be extended.

%%%%%%%%%%%%%%%%%%%%%%%%%%%%%%%%%%%%%%%%%%%%%%%%%%%%%%%%%%%%%%%
%%%%%%%%%%%%%%%%%%%%%%%%%%%%%%%%%%%%%%%%%%%%%%%%%%%%%%%%%%%%%%%
\section{Methods}
\label{sec:Methods}
In this section, we first describe how we extract the signed networks from the raw data describing the votes (Subsection \ref{sec:NetworkExtraction}). We then focus on the methods we use to partition the resulting networks (Subsection \ref{sec:PartioningMethods}).

%%%%%%%%%%%%%%%%%%%%%%%%%%%%%%%%%%%%%%%%%%%%%%%%%%%%%%%%%%%%%%%
\subsection{Network Extraction}
\label{sec:NetworkExtraction}
We first review the raw data we use in this article (Subsection \ref{sec:Dataset}), before describing the method applied to extract the signed networks (Subsection \ref{sec:Extraction}).

%%%%%%%%%%%%%%%%%%%%%%%%%%%%%%%
\subsubsection{IYP Dataset}
\label{sec:Dataset}
Like in our previous work \cite{Mendonca2015}, our row data describe the activity of the MEPs during the 7\textsuperscript{th} term of the EP, which covers the period 2009-14. They include the vote cast by each MEP for each text considered at the EP. Each MEP is also described through his name, country and political group, as well as other personal fields not used in this article. The country corresponds to one of the 28 member states (at this time) in which the MEP was elected. The political group is a transnational parliamentary coalition. The groups of the 7\textsuperscript{th} term were, by order of decreasing prevalence: \textit{EPP}: right/center-right conservatives; \textit{S\&D}: center-left; \textit{ALDE}: right/center-right neoliberals; \textit{G-EFL}: left environmentalists, progressists and regionalists; \textit{ECR}: right euroskeptics and anti-federalists; \textit{GUE-NGL}: left/far-left, socialists and communists; \textit{EFD}: right/far-right euroskeptics; NI: MEPs not belonging to any of the other groups (mainly far-right MEPs).
% \begin{itemize}
% 	\item \textit{European People's Party} (EPP): right/center-right conservatives ;
%     \item \textit{Progressive Alliance of Socialists and Democrats} (S\&D): center-left, 
%     \item \textit{Alliance of Liberals and Democrats for Europe} (ALDE): right/center-right neoliberals ;
%     \item \textit{Greens--European Free Alliance} (G-EFL): left environmentalists, progressists and regionalists ; 
%     \item \textit{European Conservatives and Reformists} (ECR): right euroskeptics and anti-federalists ;
%     \item \textit{European United Left--Nordic Green Left} (GUE-NGL): left/far-left, socialists and communists, 
%     \item \textit{Europe of Freedom and Democracy} (EFD): right/far-right euroskeptics
%     \item \textit{Non-Inscrits} (NI, non-attached members): technical group containing members not belonging to any of the other groups (during this term: mainly far-right MEPs).
% \end{itemize}
Each text is itself associated to one among $21$ specific policy domains (see \cite{Mendonca2015} for the complete list). 

In theory, these data are publicly available on the official EP website\footnote{\url{http://www.europarl.europa.eu/}}, however, in practice, accessing them is difficult. Fortunately, various institutions did all the work of compiling them, and provide them under a convenient form. The dataset we used in \cite{Mendonca2015} turned out to be incomplete: most votes from the last year of the considered term (2013-14) are missing, as well as all the amendment-related votes. For this reason, for the present work we switched to another source: the website \textit{It's Your Parliament}\footnote{\url{http://www.itsyourparliament.eu/}} (IYP), which is independently maintained by the Danish company \textit{Buhl \& Rasmussen}, and aims at both providing an easy access to these data and presenting analyses of the MEPs voting patterns.

In this dataset, one vote can be represented in $3$ ways: \textsc{For} (the MEP wants the text to be accepted), \textsc{Against} (he wants the text to be rejected) and \textsc{Abstention} (he wants to express his neutrality). It is also possible that the MEP did not vote at all, in which case he is considered as \textsc{Absent}. 

%%%%%%%%%%%%%%%%%%%%%%%%%%%%%%%
\subsubsection{Extraction Procedure}
\label{sec:Extraction}
As mentioned before, the behavior of each MEP is represented by a series of votes, corresponding to all the documents reviewed by the EP during one term. We want to extract signed networks summarizing these data. But before explaining how, we need to formalize a certain number of concepts. Let $G=(V,E,w)$ be an \textit{undirected weighted graph}, where $V$ and $E$ are the sets of vertices and edges, respectively, and $w: E \rightarrow [0;1]$ is a function associating a \textit{positive weight} to each edge. We note $n=|V|$ the number of vertices, and $w(e)$ the weight of edge $e \in E$. Consider a function $s : E \rightarrow \{+,- \}$ that assigns a \textit{sign} to each edge in $E$. An undirected weighted graph $G$ together with a function $s$ is called a \textit{signed graph}, denoted by $G = (V, E, w, s)$. An edge $e \in E$ is called negative if $s(e) = -$ and positive if $s(e) = +$. We note $E^-$ and $E^+$ the sets of negative and positive edges in a signed graph, respectively. 

The extraction method we use is similar to the one we previously proposed in \cite{Mendonca2015}, with the addition of an optional \textit{filtering step}, which makes it four-stepped. 

The first step consists in \textit{selecting} a subset of the raw data (or possibly all of it). This selection can be made along $4$ independent dimensions: policy domain (e.g. only \textit{Foreign Affairs}), political group (e.g. only S\&D), member country (e.g. only France), and time (e.g. only the year 2010-11). It is also possible to just consider all the available data over one or more dimensions, as we previously did \cite{Mendonca2015, Levorato2017}. Considering a large amount of data makes it difficult to perform a qualitative analysis of the results, and to give them an appropriate interpretation. For this reason, in this article we focus on a few countries and domains, and consider each parliamentary year separately, as explained in Section~\ref{sec:Interpretation}.

The second step consists in comparing individually all MEPs in terms of similarity of their voting behaviors, for the selected data. For this purpose, we use a variant of the vote similarity measures presented in \cite{Mendonca2015}. For a pair of MEPs $u$ and $v$ and a given text $t$, this measure noted $Sim_t(u,v)$ takes the value: $+1$ if the MEPs agree (both votes are \textsc{For} or \textsc{Against}) ; $-1$ if they disagree (one vote is \textsc{For}, the other is \textsc{Against}) ; and $0$ if at least one MEP votes \textsc{Abstention}. This measure is processed for each text and averaged to get the overall similarity between two MEPs, resulting in a real value noted $Sim(u,v)$ and ranging from $-1$ (the considered MEPs always disagree) to $+1$ (they always agree). The texts for which at least one MEP is \textsc{Absent} are not included in this average. We process the average vote similarity for every pair of \textit{active} MEPs (i.e. MEPs who voted at least once in the selected data).

The third step, which was not enforced in \cite{Mendonca2015, Levorato2017}, consists in filtering the similarity values in order to remove the ones too close to zero, which are deemed non-significant. Instead of using arbitrary thresholds, we propose an automatic method, consisting in applying $k$-means separately to the negative similarity values, and to the positive ones, with $k=2$. This allows distinguishing the values that are close to zero from the others. Indirectly, this amounts to estimating the best thresholds $\theta^-$ and $\theta^+$ such that the similarity range is split in $4$ intervals: $[-1;+1] = \{ [-1 ; \theta^-] ; ]\theta^- ; 0[ ; [0 ; \theta^+[ ; [\theta^+ ; +1] \}$. The filtering is performed by setting all values in $\{ ]\theta^- ; 0[ ; [0 ; \theta^+[ \}$ (i.e. both central intervals) to zero.

The fourth and final step allows to build the signed network based on the remaining similarity values. It consists in creating a vertex to represent each \textit{active} MEP, and connecting by an edge all pairs of MEPs whose similarity is non-zero. The sign and weight of an edge $e$ connecting two vertices $u$ and $v$ are defined as the absolute value and sign of their similarity, respectively: $w(e) = |Sim(u,v)|$ and $s(e) = sgn(Sim(u,v))$. The filtering conducted at the third step amounts to suppressing the weakest links, and the produced graph is consequently not fully connected (unlike that from \cite{Mendonca2015}). If these links indeed correspond to noise, we expect their removal to both lighten the computational load and ease the interpretation of the results.

As explained in Section~\ref{sec:Dataset}, it is possible to play with the $4$ dimensions present in our raw data (domain, group, country, and time) and used during the selection step, to obtain a number of different signed networks. There are $21$ policy domains, and we also consider all documents independently from their domain. The term is $5$-year long (2009-2014), and we consider each year separately as well as the whole term. There are $28$ countries and $8$ political groups, and in both cases we consider MEPs from specific countries/groups separately, as well as all MEPs at once. In theory, this amounts to a total of $4884$ different modalities. In practice, some of them are not usable (e.g. some groups are not represented in certain countries, resulting in an empty data selection after the first step), so our dataset contains $4150$ instances of signed networks.

%%%%%%%%%%%%%%%%%%%%%%%%%%%%%%%%%%%%%%%%%%%%%%%%%%%%%%%%%%%%%%%
\subsection{Graph Partitioning}
\label{sec:PartioningMethods}
We briefly describe the two partitioning methods, able of handling signed graphs, which we selected to analyze our signed networks.

Let us first formalize a few additional concepts. A \textit{graph partition} $P = \{C_1,...,C_k\}$ ($1 \leq k \leq n$) is a $k$-partition of $V$, i.e. a division of $V$ into $k$ non-overlapping and non-empty subsets $C_i$ ($1 \leq i \leq k$) called \textit{clusters}. For $\sigma \in \{+,-\}$, the set of positive or negative edges (depending on $\sigma$) connecting two clusters $C_i, C_j \in P$ ($1 \leq i,j \leq k$) is $E^\sigma[C_i:C_j] = \{(u,v) \in E^\sigma \mid u \in C_i, v\in C_j \},$ and its total weight is given by $\Omega^\sigma(C_i,C_j) = \sum_{e\in E^\sigma[C_i:C_j]} w(e)$.

One appropriate way of studying the structural balance of a signed network $G=(V,E,w,s)$ is by solving the \textit{Correlation Clustering} problem (CC). In its original version \cite{Bansal2002}, and consistently with the definition of structural balance given earlier in Section \ref{intro}, it consists in finding a partition of the set of vertices $V$ which maximizes both the number of positive links located \textit{inside} the clusters, and that of negative links located \textit{between} them. A relaxed version called \textit{Relaxed Correlation Clustering} problem (RCC) \cite{Doreian2009,Brusco2010} consists in searching a partition of $V$ into \textit{at most} $k$ clusters, while allowing special patterns of relationships originally considered as violations of the structural balance \cite{Doreian2009}. Both problems are formally described next.

%The sets of \textit{cut edges} and \textit{uncut edges} with respect to this partition are defined as $\cup_{1\leq i\neq j\leq l} E[C_i:C_j]$ and $\cup_{1\leq i\leq l} E[C_i:C_i]$, respectively. 
The \textit{Imbalance} $I(P)$ of a partition $P$ is defined as the total weight of positive edges located between clusters, and negative edges located inside them, i.e.
\begin{equation}
	I(P) = \sum_{1\leq i\leq k} \Omega^-(C_i,C_i) +  \sum_{1\leq i\neq j\leq k} 	\Omega^+(C_i,C_j).
	\label{IP}
\end{equation}

%CC-problem or structural balance partitioning
\begin{prblm}[CC problem]
	For a signed graph $G=(V,E,w,s)$, the \textit{Correlation Clustering problem} consists in finding a partition $P$ of $V$ such that the imbalance $I(P)$ is minimized.
\end{prblm}

In~\cite{Doreian2009}, the definition of a structurally balanced signed graph was extended in order to include relevant processes (polarization, mediation, differential popularity and subgroup internal hostility) that are counted in Eq. (\ref{IP}) as violations of the structural balance. According to this new definition, a signed graph is considered \textit{relaxed $k$-balanced} if it can be $l$-partitioned, with $l \leq k$, in such a way that: 1) all the edges within a given cluster have the same sign (not necessarily $+$) ; and 2) all the edges between two given clusters have the same sign (not necessarily $-$).

Using this new definition, the structural balance was generalized to a version named \textit{Relaxed Structural Balance}~\cite{Doreian2009}, resulting in a new definition for the imbalance of a graph partition. For an $l$-partition $P=\{C_1,...,C_l\}$, the \textit{Relaxed Imbalance} $RI(P)$ is defined as
\begin{align}
	RI(P) = &\sum_{1 \leq i \leq l} \min\{\Omega^+(C_i,C_i),\Omega^-(C_i,C_i)\} \\
	&+\sum_{1 \leq i \neq j \leq l} \min\{\Omega^+(C_i,C_j),\Omega^-(C_i,C_j)\}.
	\label{RIP}
\end{align}

This generalized imbalance defines a new criteria to evaluate balancing in a signed graph, and gives rise to the following graph clustering problem.

%Relaxed CC-problem or relaxed structural balance partitioning
\begin{prblm}[RCC problem]
	Let $G=(V,E,w,s)$ be a signed graph, and $k$ an integer value satisfying $1 \leq k \leq n$. The \textit{Relaxed Correlation Clustering problem} consists in finding an $l$-partition $P$ of $V$, with $l \leq k$, such that the relaxed imbalance $RI(P)$ is minimized.
\end{prblm}

It is worth noticing that, for a given graph, the RCC solution is necessarily equally or more balanced than the CC one.
%: in the worst case, the CC solution holds, and in the best case, the graph can be partitioned more efficiently by taking advantage of the fact some violations of the original structural balance are accepted in its relaxation.

Both CC and RCC problems have been proved to be $NP$-hard \cite{Bansal2002, Figueiredo2013}. Integer programming-based methods can be used to solve both problems to optimality \cite{Ales2016, Figueiredo2013}, and numerical experiments have shown that the additional parameter $k$, in the RCC definition, turns the graph clustering problem more difficult to solve numerically \cite{Figueiredo2013}. Both problems can be efficiently solved by the \textit{Iterated Local Search} (ILS) procedures described in \cite{Levorato2017a}, namely ILS-CC and ILS-RCC. 
%ILS \cite{Lourenco2003} belongs to a class of local search methods proposed in the literature to efficiently investigate the feasible solution space of combinatorial problems. Broadly speaking, ILS is a metaheuristic based on stochastic multi-restart searches, which iteratively applies an \textit{ad hoc} local search to perturbations of the current best solution, generating a sequence of locally optimal solutions. The ILS-CC and ILS-RCC methods are fully described in \cite{Levorato2017a}. 
As a baseline, we also use the exact algorithm described in \cite{Ales2016} to solve CC, which we call Ex-CC in the rest of the article. 

We apply both ILS procedures in order to identify structural balanced partitions in each extracted signed network. The following strategy was adopted to set the parameter $k$, which is an input of the RCC problem. First, we apply ILS-CC and obtain an optimal partition containing $k'$ clusters. Second, we apply ILS-RCC to the same network with parameter $k \in \{k', k'+1, k'+2\}$, producing $3$ different network partitions. We consider and compare all obtained partitions when interpreting our results relatively to the studied system \todoNA{Actually, we have used $k$ and $k+1$, so omitted $k+2$}\todoVL{I don't think so, see italy in figure 6}.

In the experimental part, we compare partitions using the \textit{Normalized Mutual Information} (NMI), a measure widespread in the fields of clustering \cite{Strehl2002} and community detection \cite{Fortunato2010}. It ranges from $0$ (the partitions are completely different) to $1$ (they are exactly similar). This measure takes into account the possible hierarchical relations between the considered partitions. For instance, if one cluster from the first partition corresponds to two distinct clusters in the second partition, the penalty will be lower than if there is no match at all \cite{Danon2005}.

\section{Experimental Assessment}
\label{sec:Stats}
In this section, we first study the effect of the filtering step of our extraction process on both the obtained networks and the partitioning algorithms (Subsection~\ref{sec:FilteringEffect}). We then assess the quality of the heuristic previously proposed to solve the CC problem (Subsection~\ref{sec:HeuristicEvaluation}). Note that the heuristic we use to solve RCC relies on exactly the same operators, which is why we do not need to explicitly evaluate it for this problem. For both points (filtering and heuristic), we base our analysis on the $4150$ signed network instances constituting our whole dataset. Both our data\footnote{\url{https://doi.org/10.6084/m9.figshare.5785833}} and source code\footnote{\url{https://github.com/CompNet/NetVotes}} are publicly available online.

For convenience, we will use the following notations: $ExCC(G)$ and $ILSCC(G)$ are the imbalance values obtained through the corresponding partitioning algorithms, expressed in total edge weight as defined by equation \ref{IP}. It is sometimes more convenient to discuss an imbalance described in terms of percent of the total link weight of the graph, in which case we add a \% to our notation. Finally, the filtered version of a graph $G$ is noted $G_f$.

%%%%%%%%%%%%%%%%%%%%%%%%%%%%%%%%%%%%%%%%%%%%%%%%%%%%%%%%%%%%%%%
\subsection{Effect of the Filtering Step}
\label{sec:FilteringEffect}
As explained in Section~\ref{sec:NetworkExtraction}, we added a filtering step to our previously defined extraction process \cite{Mendonca2015}: the goal of this subsection is to assess its influence on both the resulting graphs (Subsection~\ref{sec:FilteringEffectGraphs}) and the partitioning algorithms (Subsection~\ref{sec:FilteringEffectAlgorithms}).

%%%%%%%%%%%%%%%%%%%%%%%%%%%%%
\subsubsection{Effect on the Graphs}
\label{sec:FilteringEffectGraphs}
\begin{figure*}[htb!]
	\centering
	\includegraphics[width=0.45\textwidth]{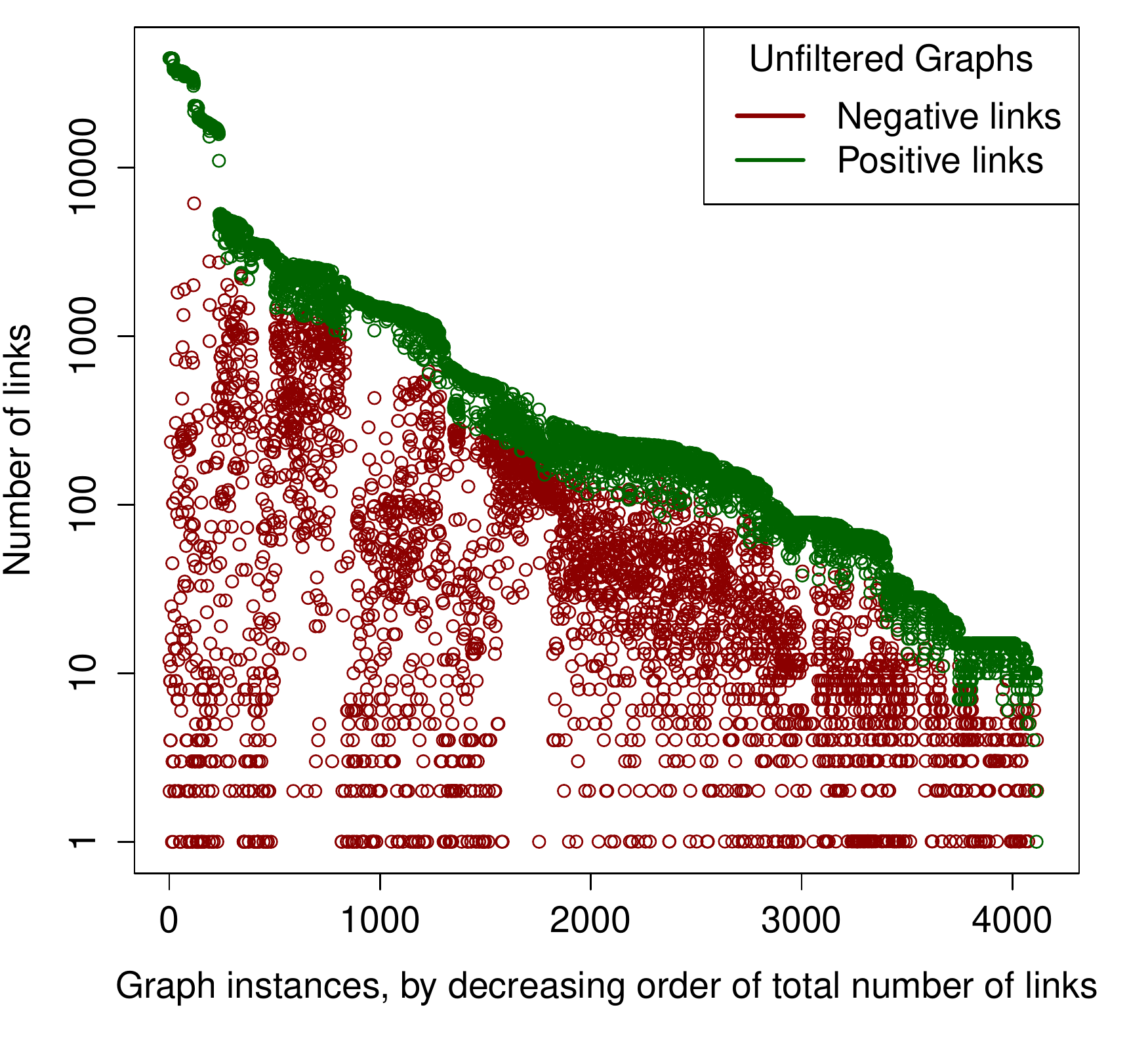}
	\includegraphics[width=0.45\textwidth]{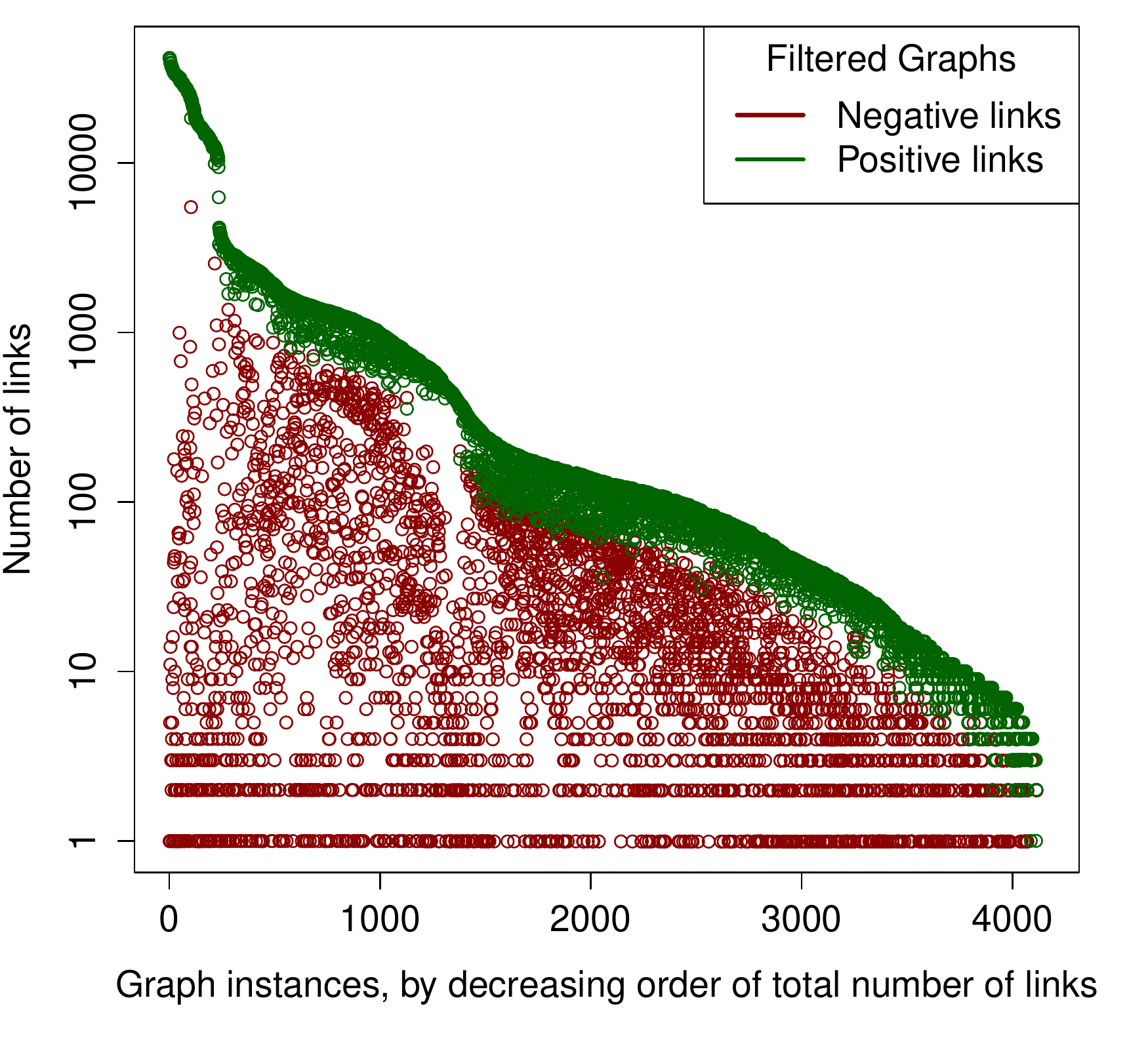}
    \vspace{-0.4cm}
	\caption{Numbers of positive (green) and negative (red) edges in all unfiltered (left) and filtered (right) networks.}
	\label{fig:LinkCounts}
\end{figure*}

The plots from Figure~\ref{fig:LinkCounts} show the numbers of positive (in green) and negative (in red) links for each network in the dataset, \textit{without} (left-hand plot) and \textit{with} filtering (right-hand plot). On the $x$ axis, the networks are ordered by decreasing number of links. The $y$ axis uses a logarithmic scale for readability matters. There obviously is a decrease in the numbers of positive and negative links when filtering the graphs ($43\%$ of the links are removed, in average). However it is worth noticing that the general distribution does not change (if anything, it is smoothed) and the proportions of positive and negative links are also preserved. This is even truer when considering weights instead of link counts (not plotted here), since the filtering removes the weaker links (only $26\%$ of the network weight, in average). This point is important, because the considered partitioning methods take weights into account.

Another important point besides link distribution is graph connectivity. We focus on the giant component of the unfiltered graphs (which by definition are almost completely connected) and study how they are affected by the filtering step. For the overwhelming majority of the instances ($66\%$), filtering does not make the graph disconnected. For $23\%$ (resp. $8\%$) of them, it splits them in $2$ (resp. $3$) components. But in this case, the larger component represent $87\%$ (resp. $76\%$) of the original component, which means we still have one giant component. Filtering splits the rest of the graphs into up to $10$ components, but this represents a very small proportion of the dataset ($3\%$), constituted of very small graphs, and the giant component is also preserved in these cases.

%%%%%%%%%%%%%%%%%%%%%%%%%%%%%
\subsubsection{Effect on the Partitioning Algorithms}
\label{sec:FilteringEffectAlgorithms}
According to our experiments, the filtering step generally leads to a decrease in terms of partitioning processing time. This effect is more pronounced for graphs whose unfiltered partitioning requires more time, since faster ones offer little room for improvement. For Ex-CC, filtering cuts the processing time by $5$--$10\%$ in those graphs, whereas for ILS-CC it is close to $90\%$.
We now turn to the quality of the detected clusters. We focus on Ex-CC and ignore ILS-CC, because only the analysis of optimal solutions can allow us to assess the true effect of filtering. We first consider the numbers of clusters that this algorithm detects. For almost all instances, we observe an increase in the number of clusters in the filtered graphs. But for a few of them, there is a decrease: this is explained by the fact that removing an edge can decrease cluster cohesion, if this edge is positive, but also cluster separation, if it is negative. We have no ground truth (group membership is not a reference, as shown in Section \ref{sec:Interpretation}), so it is difficult to objectively assess whether an increase in the number of clusters is a good thing or not. However, we can use our knowledge of the data: since they represent voting behaviors, we expect $1$ cluster if there is a consensus, $2$ if there is a clear \textsc{For}/\textsc{Against} opposition, and a very few more if some MEPs swing in-between. 
We observe that on the unfiltered networks Ex-CC detects a single cluster for most networks ($55\%$ of them). It seems unlikely that a consensus would emerge that often. After filtering, this proportion drops to $38\%$, which seems more reasonable.

\begin{figure*}[htb!]
	\centering
	\includegraphics[width=0.45\textwidth]{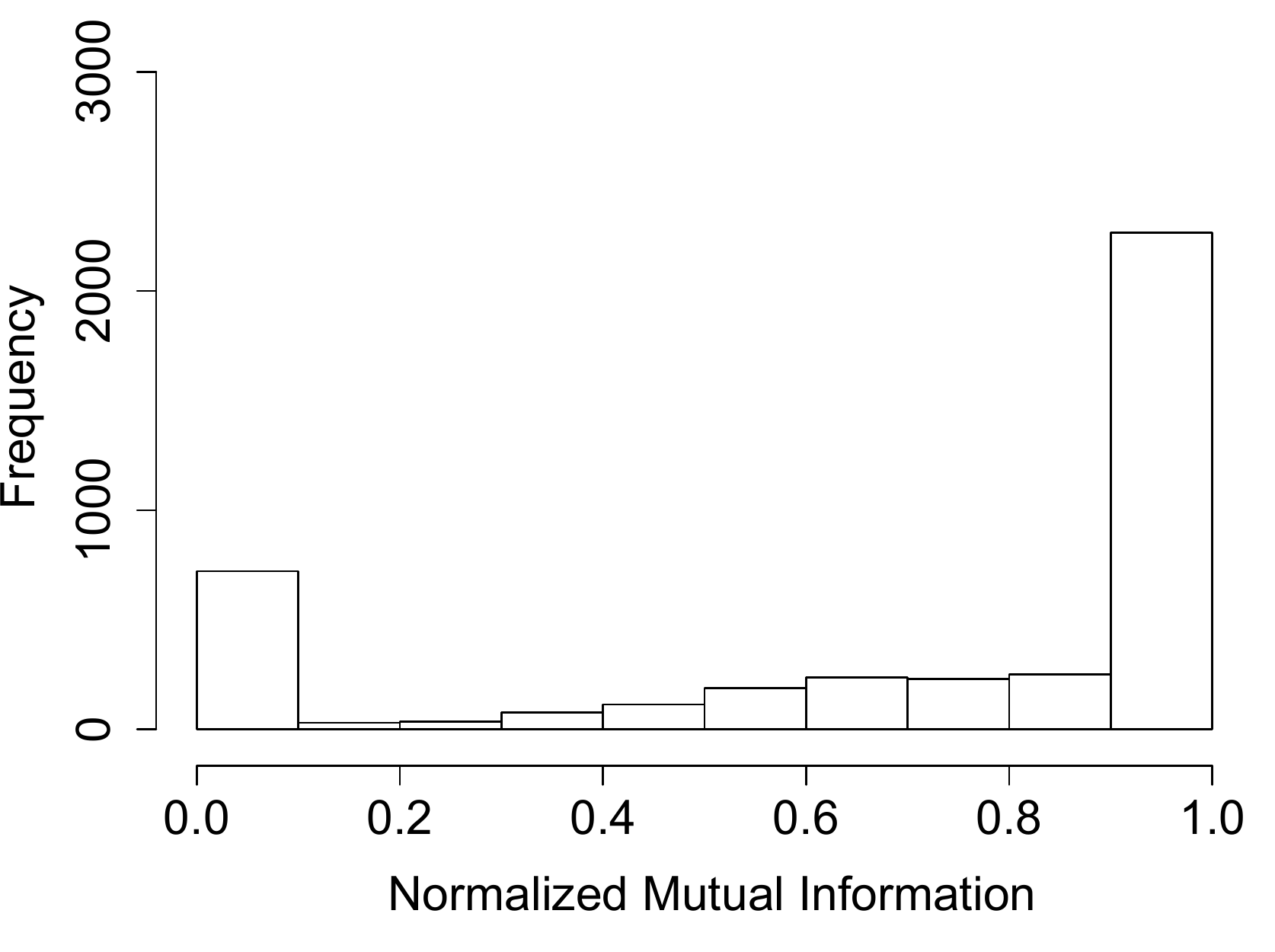}
    \includegraphics[width=0.45\textwidth]{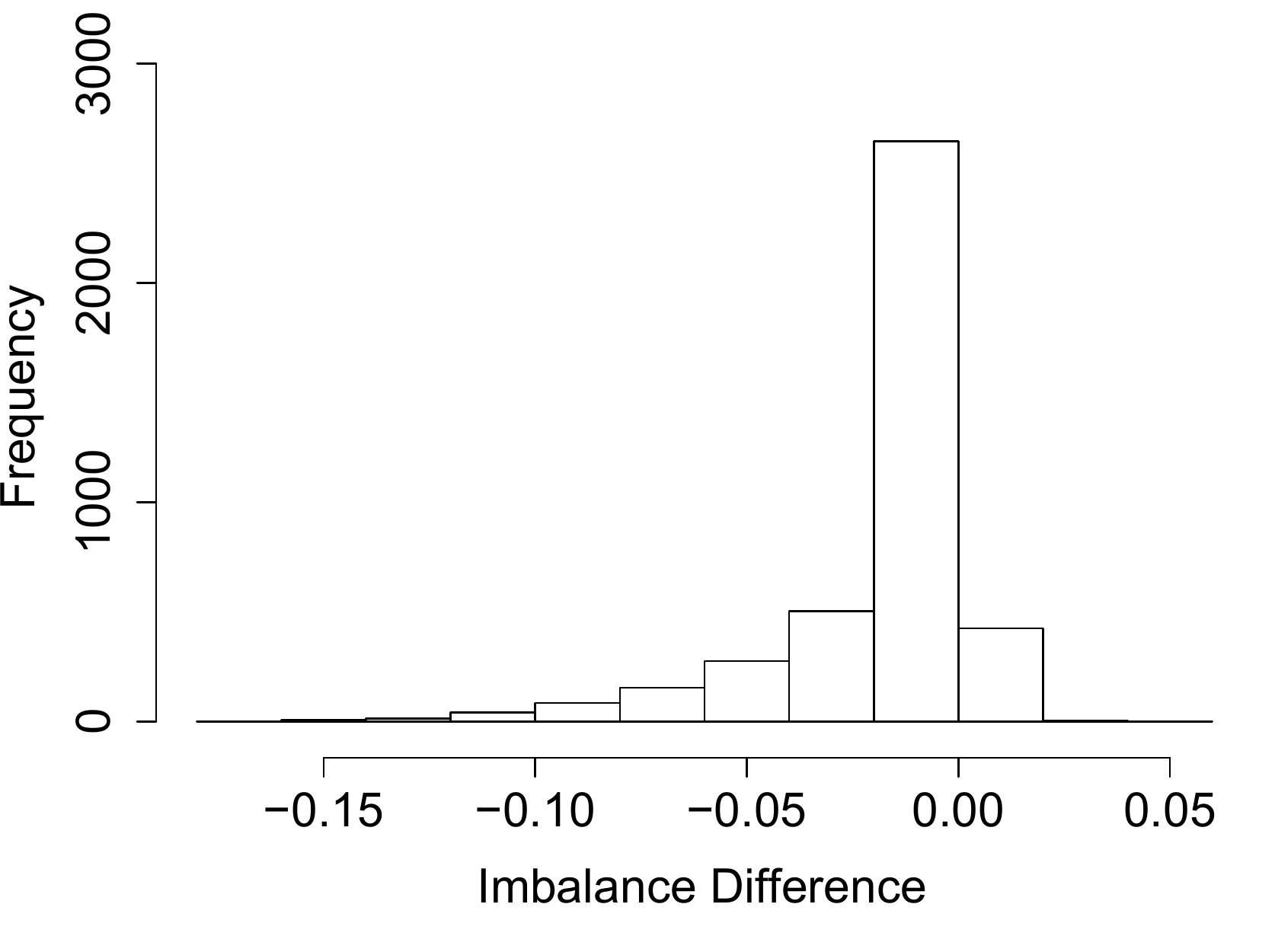}
    \vspace{-0.4cm}
	\caption{Distribution of the NMI (left) and of imbalance difference (center) between partitions detected by Ex-CC \textit{with} vs. \textit{without} filtering. On the right: execution time of Ex-CC and ILS-CC as a function of the network size.}
	\label{fig:FilterNMI_Imb}
\end{figure*}

But two partitions can contain the same number of clusters while being very different (and oppositely: contain relatively different numbers of clusters, but be hierarchically related). It is thus necessary to assess how similar two solutions on an unfiltered graph and its filtered counterpart are. For this purpose, we use the NMI (cf. Subsection \ref{sec:PartioningMethods}), which is showed in the left plot of Figure~\ref{fig:FilterNMI_Imb}. The NMI is close to $1$ for most instances ($\geq 0.8$ for $61\%$ of them), meaning the identified partitions are not affected much by the filtering, despite the previously observed variation in the number of clusters. 

We now turn to the quality of the partitions expressed in terms of imbalance. The center plot in Figure~\ref{fig:FilterNMI_Imb} displays the distribution of $ExCC\%(G)-ExCC\%(G_f)$,
i.e. the difference of percent imbalance between the partitions detected on an unfiltered graph vs. the corresponding filtered graph. The values are distributed around $0$, meaning the quality in terms of imbalance does not change much when filtering. Based on the previous observations, we can conclude the filtering step only marginally affects the partitions detected by Ex-CC.

%%%%%%%%%%%%%%%%%%%%%%%%%%%%%%%%%%%%%%%%%%%%%%%%%%%%%%%%%%%%%%%
\subsection{Evaluation of the Heuristic}
\label{sec:HeuristicEvaluation}
We now turn to the evaluation of ILS-CC, the heuristic method proposed to solve the CC problem. We consider two aspects: the quality of the identified partitions, in terms of imbalance, and the gain in computational time. We use the parameter values recommended by the authors of the algorithm \cite{Levorato2017a}. Based on the conclusion of the previous subsection, we apply the algorithm to the filtered networks only.

\begin{figure}[htb!]
	\centering
    \includegraphics[width=0.90\columnwidth]{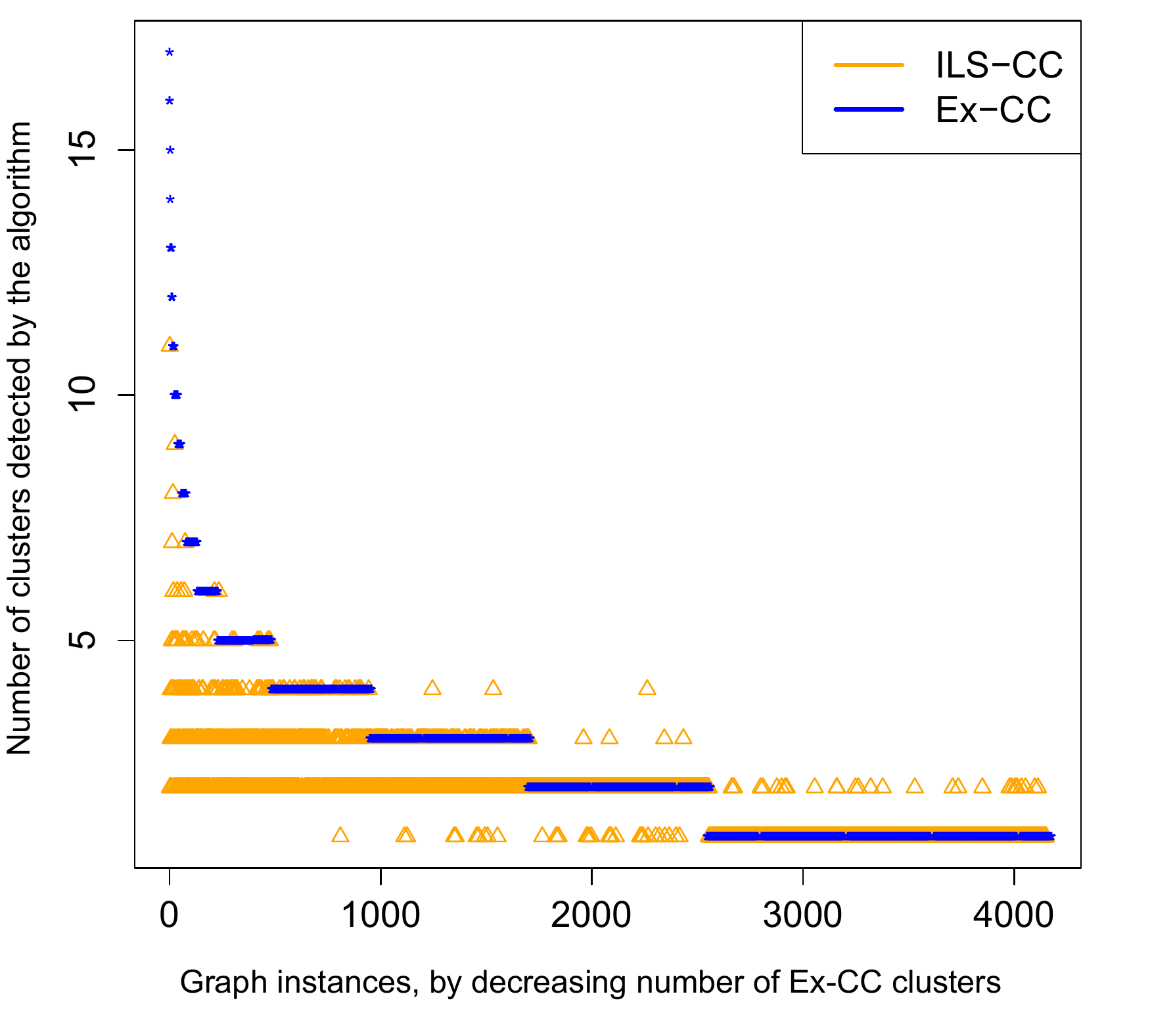}
    \vspace{-0.4cm}
	\caption{Comparison of Ex-CC and ILS-CC in terms of numbers of clusters.}
	\label{fig:ILSvsCCNbrClust}
\end{figure}

\begin{figure*}[htb!]
	\centering
    \includegraphics[width=0.45\textwidth]{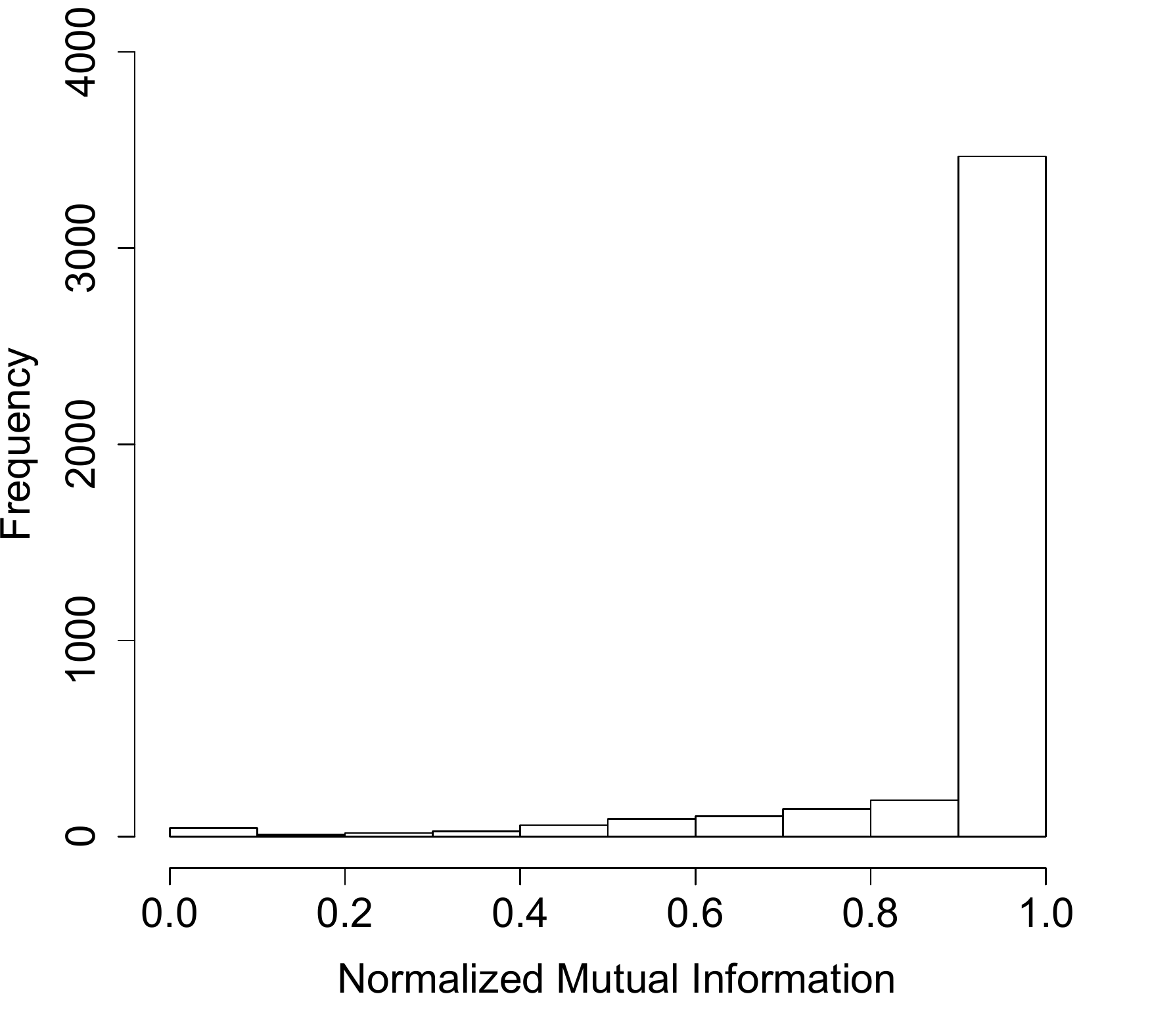}
    \includegraphics[width=0.45\textwidth]{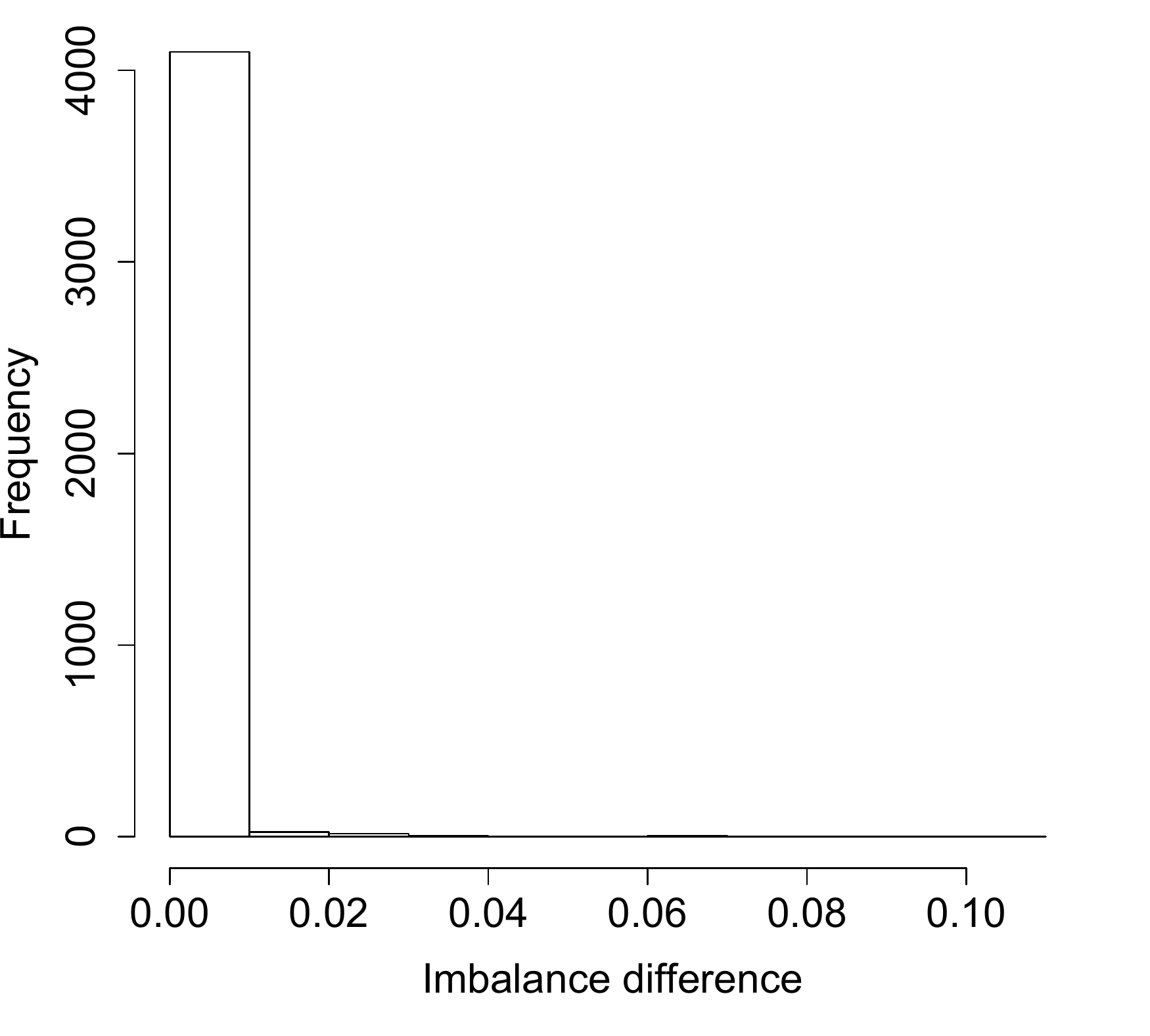}
    \vspace{-0.4cm}
	\caption{Comparison of Ex-CC and ILS-CC in terms of NMI distribution (left) and Imbalance difference (right).}
	\label{fig:ILSvsCC}
\end{figure*}

Figures~\ref{fig:ILSvsCC} and \ref{fig:ILSvsCCNbrClust} display the obtained results. The former shows the numbers of clusters detected by Ex-CC (in blue) and ILS-CC (orange). The networks are ordered by decreasing number of Ex-CC clusters. One can observe that the approximate approach gets very close to Ex-CC. In Figure~\ref{fig:ILSvsCC}, the left plot displays the NMI distribution, and we see that in $83\%$ of the cases, ILS-CC identifies a partition very similar to the optimal one ($NMI \geq .8$). The right plot is the distribution of imbalance difference $ILSCC\%(G_f) - ExCC\%(G_f)$, which shows that the results output by ILS-CC are optimal or near-optimal for almost all instances ($98\%$ of the cases). Qualitatively speaking, the heuristic leads to excellent approximations on these data.

\begin{figure}[htb!]
	\centering
     \includegraphics[width=0.90\columnwidth]{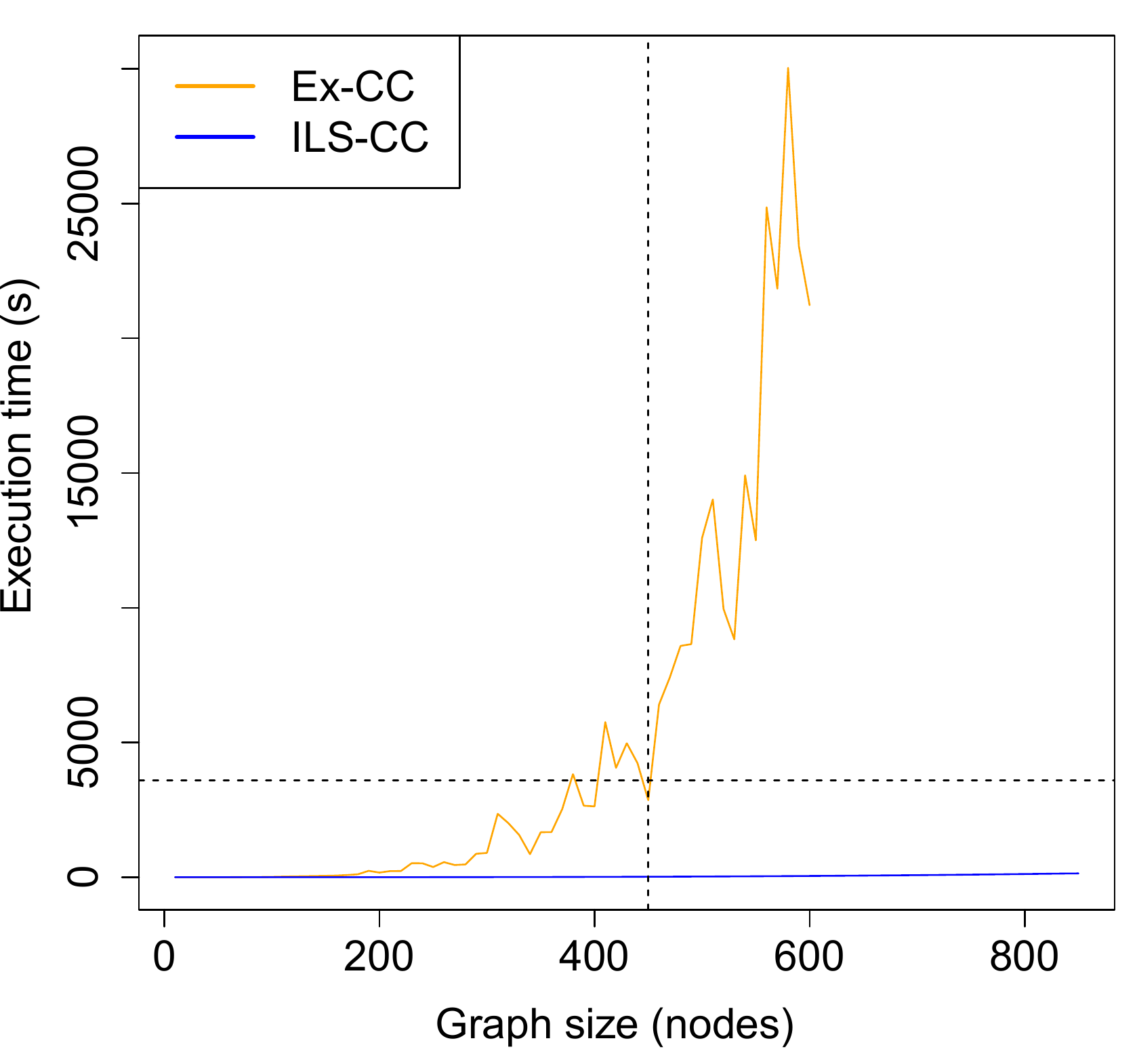}
	\caption{Execution time of Ex-CC and ILS-CC (in s), as a function of the network size (in nodes).}
     \label{fig:ILSvsCCTime}
\end{figure}

We now turn to the computational gain of the heuristic. Figure~\ref{fig:ILSvsCCTime} shows the evolution of the processing times for both Ex-CC (orange) and ILS-CC (blue), expressed in seconds, as a function of the number of nodes in the processed network. We used a 24-core AMD Opteron 2.6 GHz CPU with 512 GB RAM. The benchmark was generated randomly by sampling an increasing number of nodes from the largest unfiltered network. The obtained sizes range from $10$ to $850$ nodes, by $10$-node steps. Since consecutive graphs are independently drawn, we expect some fluctuations in the obtained durations. Our first observation is that it becomes very hard for Ex-CC to solve instances larger than $450$ vertices (vertical dashed line in the figure), as the necessary computing time passes the $1$ hour limit (shown with a horizontal dashed line). We could not finish processing networks larger than $600$ nodes, which is why the line is interrupted. On the contrary, ILS-CC is very fast even for the largest graphs: the time needed is of the order of a few minutes ($142.48$ s at most). Note that the quality of the estimated partitions are also very good, similar to what was described in the previous subsection.

%%%%%%%%%%%%%%%%%%%%%%%%%%%%%%%%%%%%%%%%%%%%%%%%%%%%%%%%%%%%%%%
%%%%%%%%%%%%%%%%%%%%%%%%%%%%%%%%%%%%%%%%%%%%%%%%%%%%%%%%%%%%%%%
\section{Interpretation of Specific Cases}
\label{sec:Interpretation}
The assessment conducted in the previous section was purely quantitative, and based on the whole dataset. In order to explore the usefulness of signed graph partitioning in a more qualitative way, we now focus on a few specific cases of interest. We discuss the results obtained while solving both CC and RCC on the networks representing French and Italian MEPs, for $2$ policy domains: \textit{Agriculture \& Rural Development} (AGRI) and \textit{Economic \& Monetary Affairs} (ECON).
%, and \textit{Women's Right \& Gender Equality} (FEMM). 
We chose those topics because of their potentially polarizing nature: AGRI because the \textit{Common Agricultural Policy} (CAP) has historically been of utmost importance for France due to the prevalence of the agricultural sector in its economy, but a part of the population wants to leave the industrial model of production; and ECON because the subprime mortgage crisis started just before the considered term.
%; and FEMM because this topic is likely to highlight strong ideological disagreements. 
We selected France (FR) because of our knowledge of its politics, whereas Italy (IT) is interesting as a reference, since it has roughly the same number of MEPs, and a relatively close culture. Moreover, for each case, we focus on the year for which the results are the most illustrative.

In the rest of this section, we exhibit two types of graphs: individual vs. cluster networks. In an \textit{individual network} (Figures~\ref{fig:FrAgri}a,e and \ref{fig:FrEcon}a,e), vertices represent MEPs and edges represent filtered vote similarity values. We use a circular layout, and gather MEPs by political group. Each group is characterized by a color: GUE-NGL (red), G-ELF (green), S\&D (pink), ALDE (orange), EPP (light blue), ECR (dark blue), EFD (purple) and NI (brown). In a \textit{cluster network} (Figures~\ref{fig:FrAgri}b-d,f-g and \ref{fig:FrEcon}b-d,f-g), each vertex represents a cluster, and each positive (resp. negative) edge corresponds to the set of all individual positive (resp. negative) edges between the MEPs constituting these clusters. For each cluster, we indicate the groups of the MEPs constituting it, and its proportions of internal positive and negative links, relatively to the total network weight. The cluster itself is represented by a pie chart also reflecting these proportions. Similar proportions are provided also for each edge. Finally, in both types of graphs, positive (resp. negative) edges are drawn in green (resp. red).

%%%%%%%%%%%%%%%%%%%%%%%%%%%%%%%%%%%%%%%%%%%%%%%%%%%%%%%%%%%%%%%
\subsection{Agriculture \& Rural Development}
\label{sec:ResultsAgri}

\begin{figure*}[htb!]
	\centering
	\includegraphics[width=0.21\textwidth]{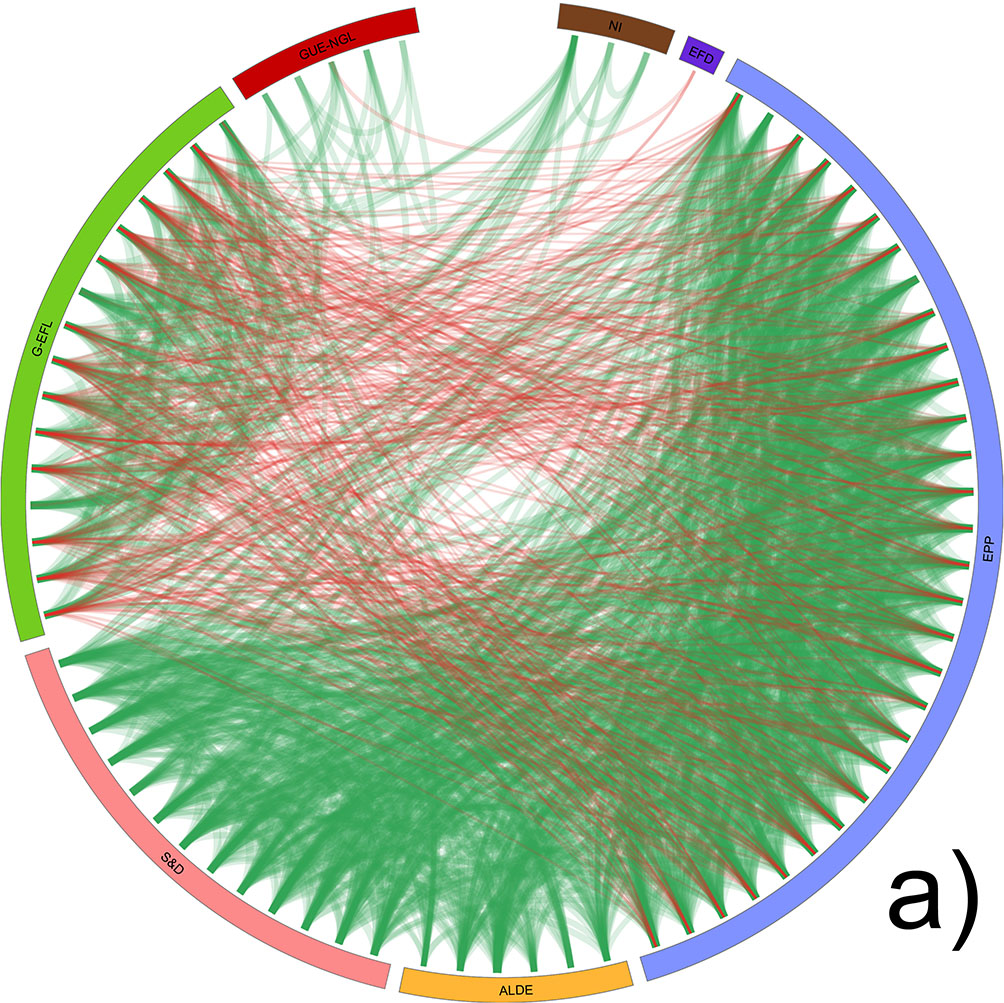}
    \includegraphics[width=0.25\textwidth]{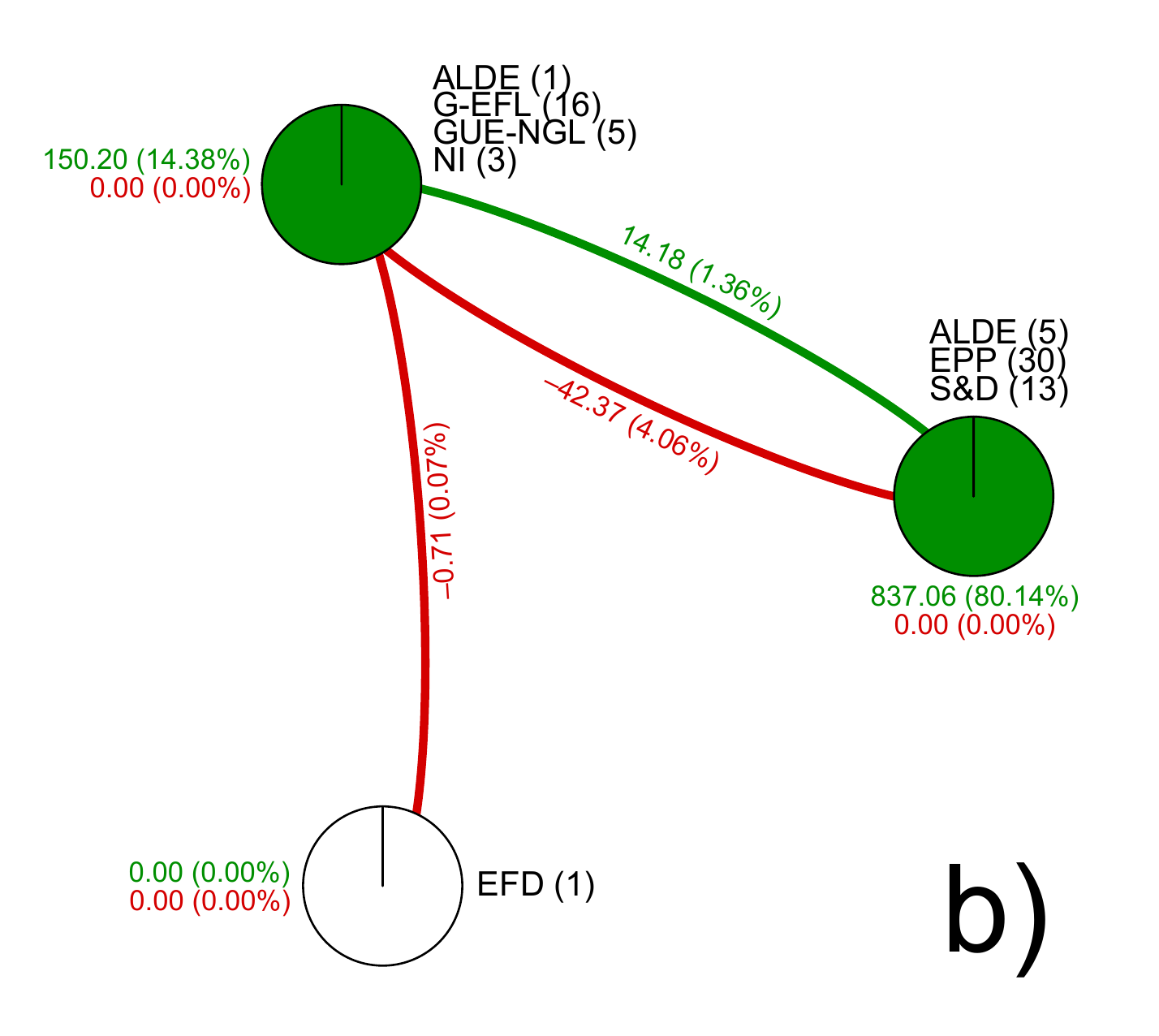}
    \includegraphics[width=0.25\textwidth]{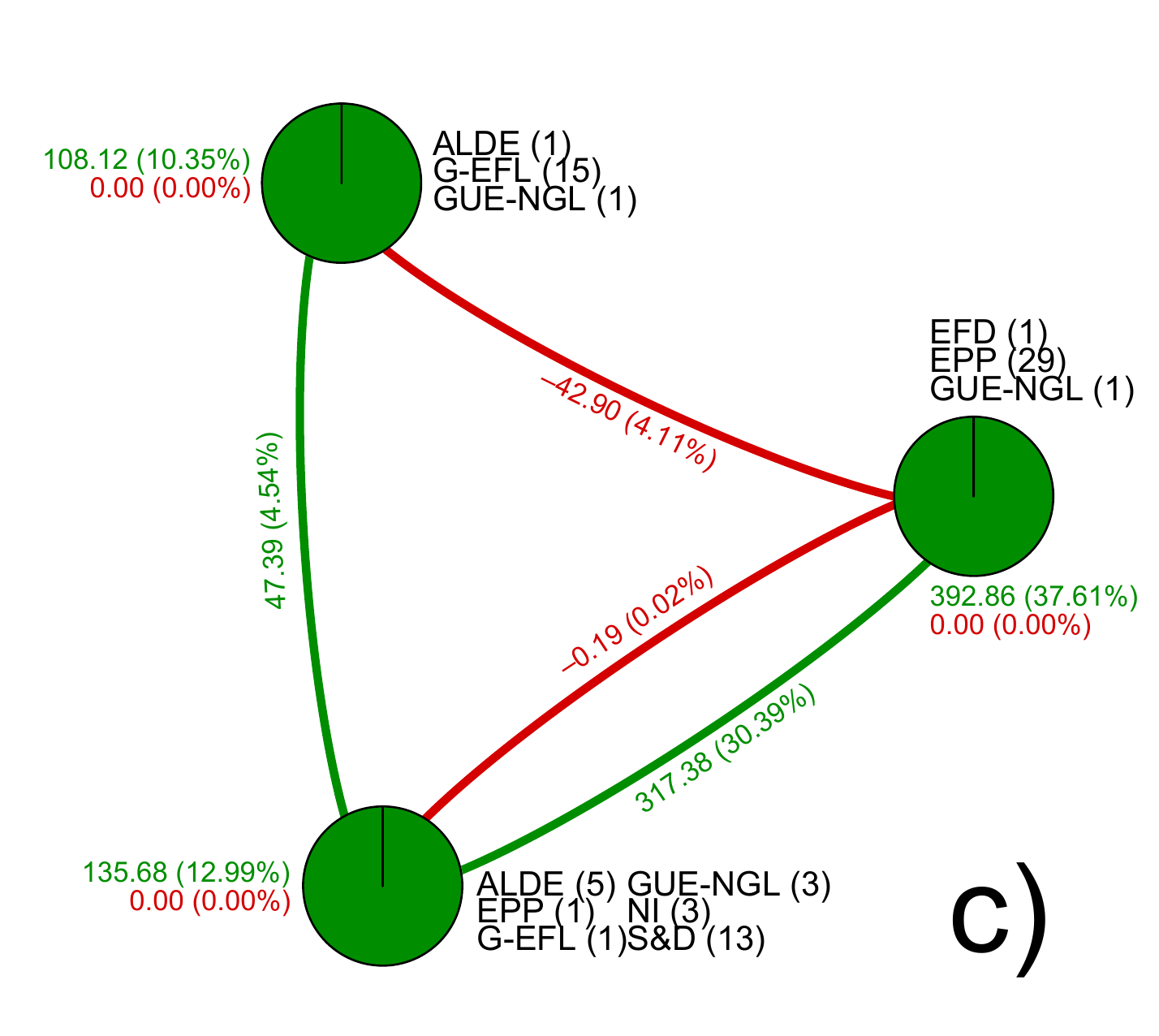}
    \includegraphics[width=0.25\textwidth]{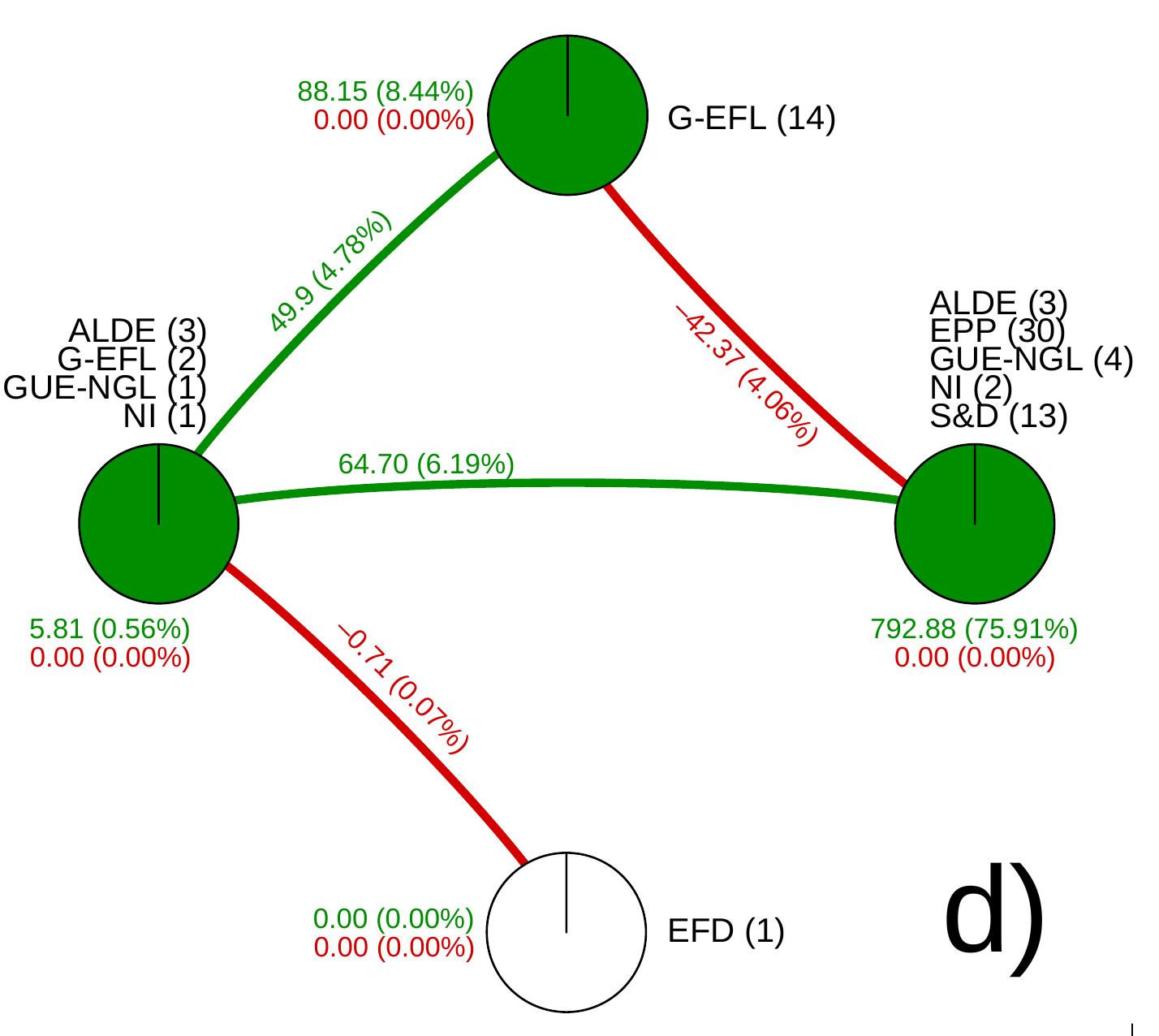}\\
	\includegraphics[width=0.21\textwidth]{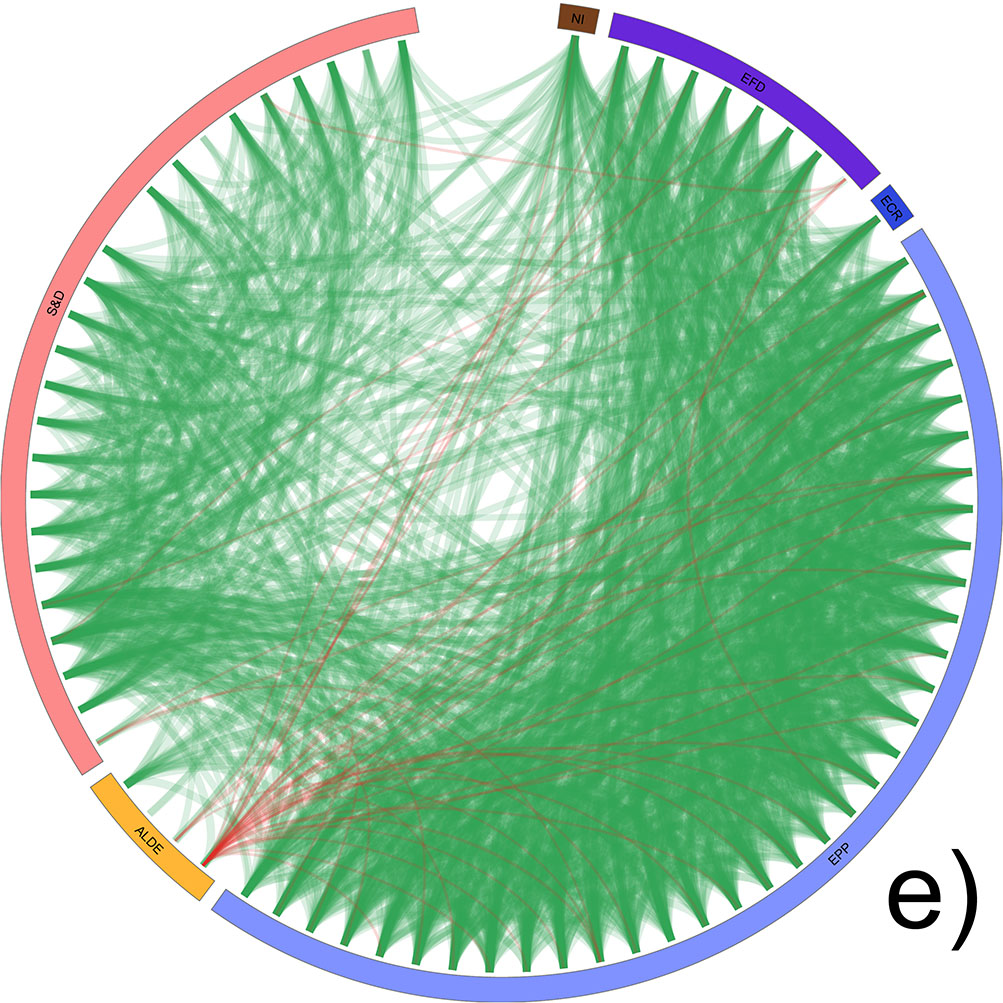}
    \includegraphics[width=0.25\textwidth]{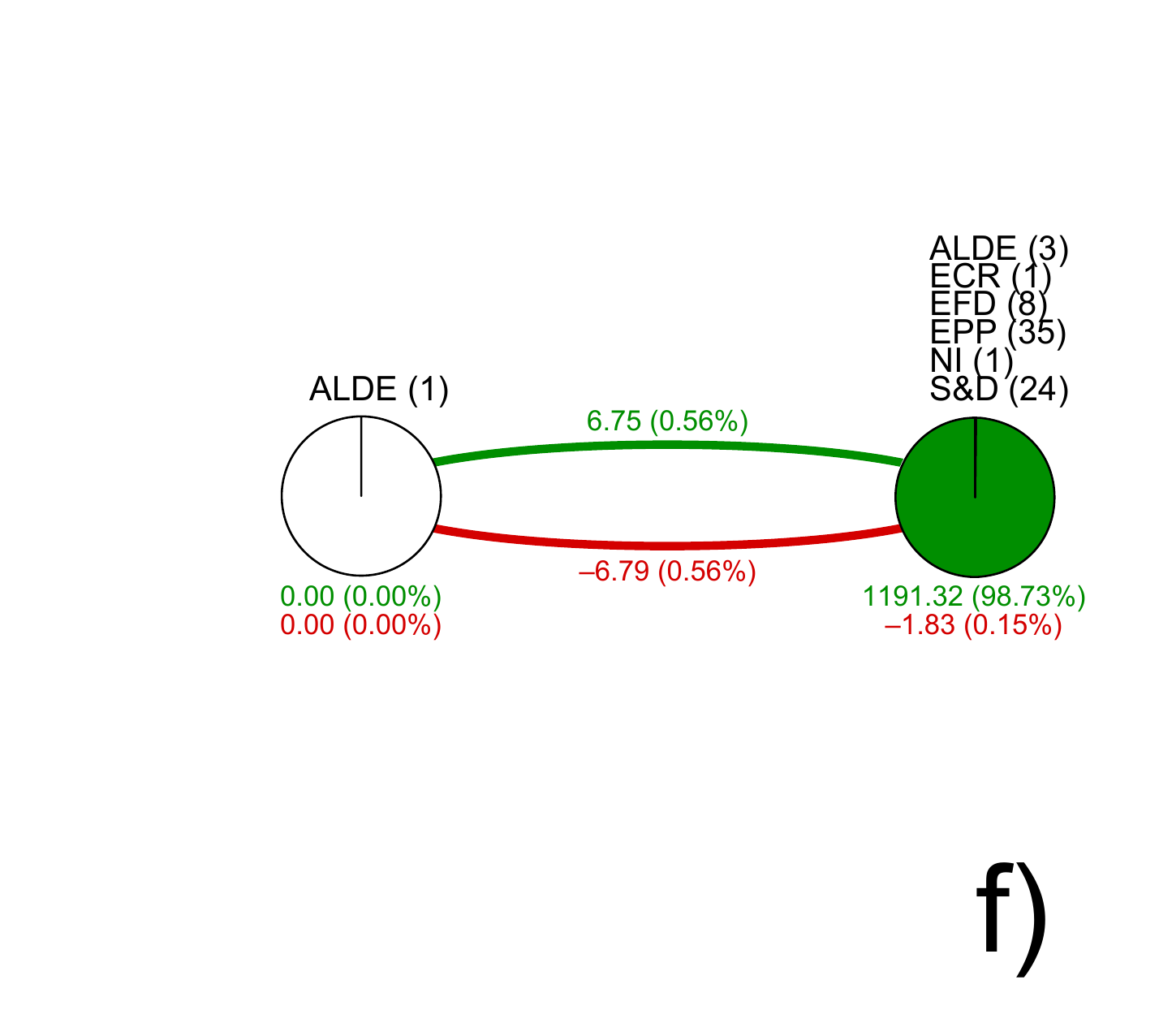}
    \includegraphics[width=0.25\textwidth]{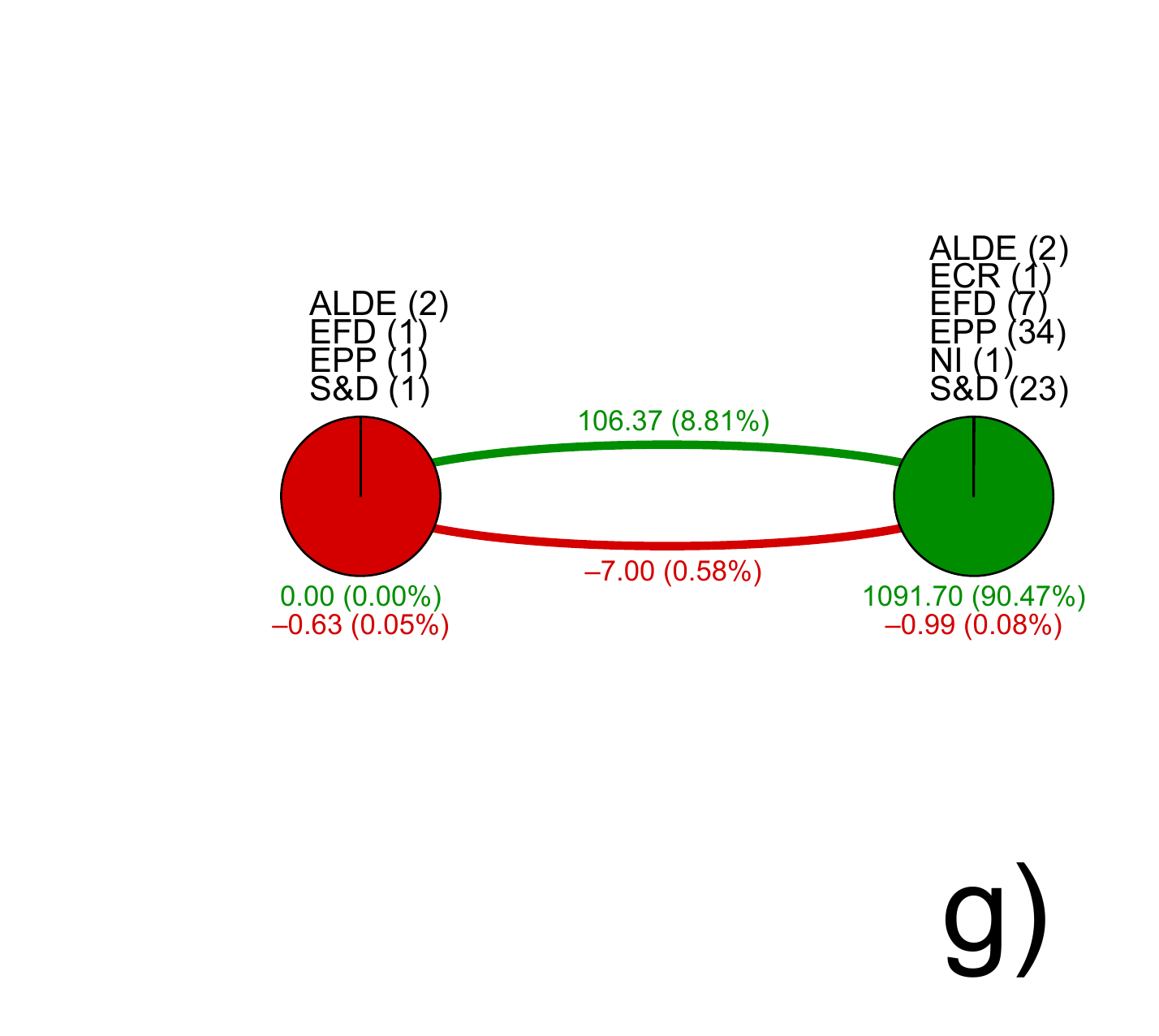}
    \includegraphics[width=0.25\textwidth]{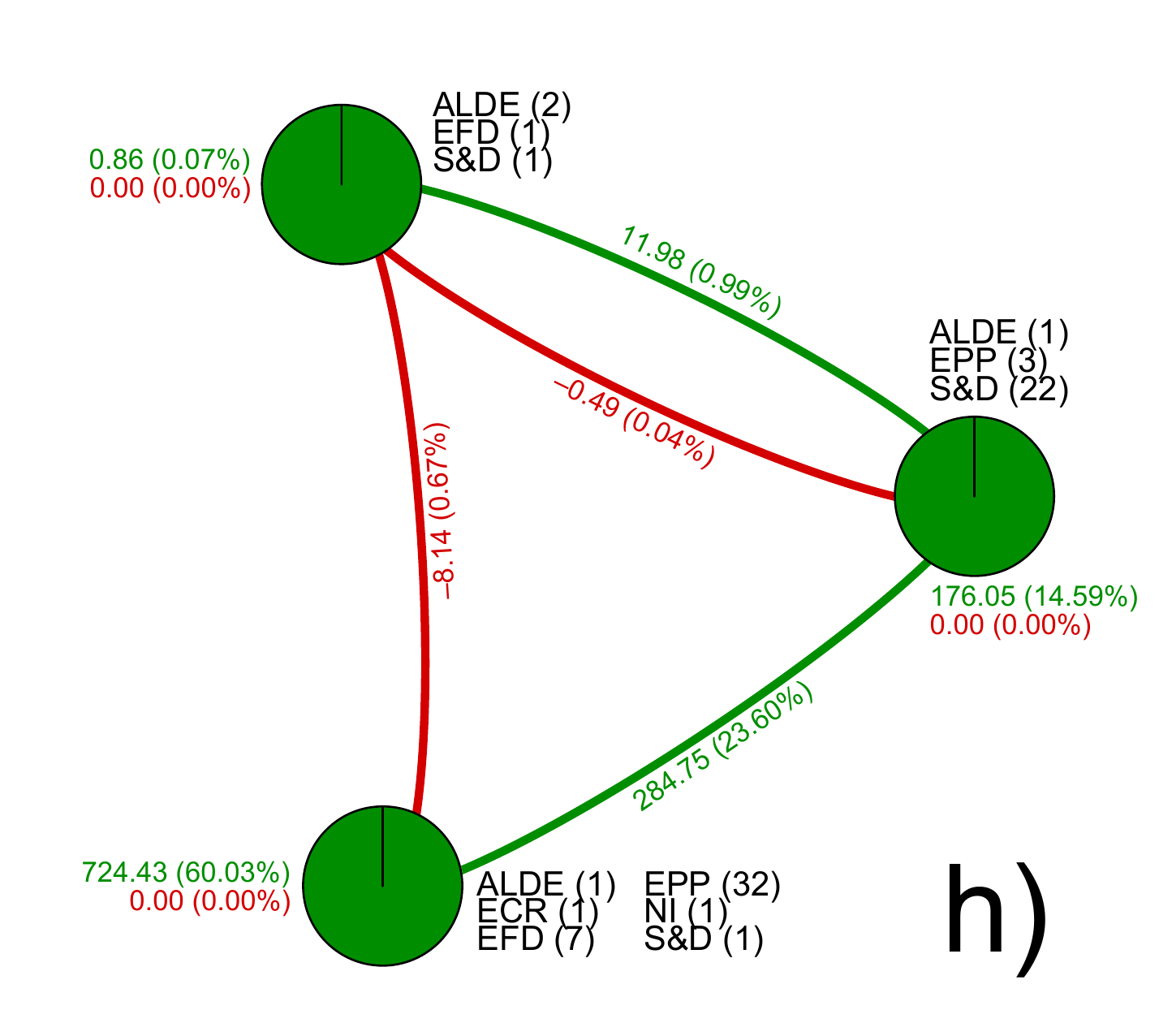}
	\caption{French (top) and Italian (bottom) MEPs for the AGRI domain during the year 2012-13: individual networks (a,e) ; and cluster networks for the CC problem (b,f), the RCC problem with $k=2$ (g), $k=3$ (c,h), and $k=4$ (d).}
	\label{fig:FrAgri} 
\end{figure*}

For AGRI, we focus on the year 2012-13. The top plots in Figure\ref{fig:FrAgri} represent the network of the French MEPs, and $3$ different partitions estimated for this network. The first partition (b) is the optimal solution to CC, which contains $3$ clusters and has an imbalance of $14.18$ (1.35\% of the total weight). There are two large clusters of similar size: the left one is largely dominated by the environmentalists (G-EFL) and also contains the radical left (GUE-NGL) and NI ; and the right cluster contains the center-left (S\&D), right and center-right (EPP and ALDE) groups. Both clusters have a majority of positive internal and negative external links. The third cluster (at the bottom) is a single node corresponding to \textit{Philippe de Villiers}, the only French member of the right-wing euroskeptic EFD group. Unlike the three members of the other euroskeptic group (NI), he is connected to the rest of the graph only by negative links. This partition displays a clear left/right divide, with the exception of the three NI members, who are put together with the left/environmentalists. This divide can be explained when considering the texts voted this year, among which many concern animal rights and related matters (questions of great importance for these groups). The fact one ALDE member was put with the Greens supports this, since it corresponds to \textit{Corrine Lepage}, former Minister of the Environment in a right-wing French government. One could expect the S\&D to vote similarly to the rest of the left on these topics. However, even more texts voted during this year concern the CAP, on which the center-left is more likely to side with the right. This also explains the position of NI, which is more likely to opportunistically support resolutions in favor of small family-owned farms (and therefore vote like the radical left).

The other partitions are solutions of the RCC problem, with various $k$ values. For $k=3$ (we start from the optimal number of clusters for CC, as explained in Section~\ref{sec:PartioningMethods}), represented in plot (c), we get a lower imbalance than with CC ($0.19$), which is to be expected. The partition differs from the first one in that it identifies $3$ clusters of comparable sizes (no more singleton cluster). Both large clusters from the previous partition loose a number of members, which are gathered to form a new, intermediary group. It contains some of the radical left (GUE-NGL), the center-left (S\&D), center-right (ALDE) as well as the NI group. The two other clusters are the environmentalists (G-EFL with Lepage) and the rest of the right (EPP with de Villiers), respectively. This partition is interesting, because it manages to identify a cluster of moderate MEPs, which sometimes vote like the environmentalists, and sometimes like the right. The cluster graph consequently takes the form of an imbalanced (in terms of CC) triangle in which the environmentalists and the right are in opposition, whereas the moderate are positively connected to both. This type of structure could be identified only thanks to the relaxed nature of RCC, which allows here to have positive links \textit{between} clusters. 

The $k=4$ partition is quite similar to the CC partition, in the sense S\&D and EPP are in the same cluster, and de Villiers is apart again. But NI and GUE-NGL are now part of the EPP-dominated cluster, and there is a fourth cluster formed by MEPs from almost all political groups except EPP. Due to this heterogeneity, this last cluster is very difficult to interpret. Yet, the partition reaches a perfect zero imbalance, which means absolutely no link is misplaced. This highlights the fact increasing $k$ will decrease the imbalance, but not necessarily make the partition more informative, from the application point of view. This means that the problem, as it is formulated, does not allow to completely automate the process of identifying the best partition for the end-user.

%\FloatBarrier
\begin{figure*}[htb!]
	\centering
	\includegraphics[width=0.21\textwidth]{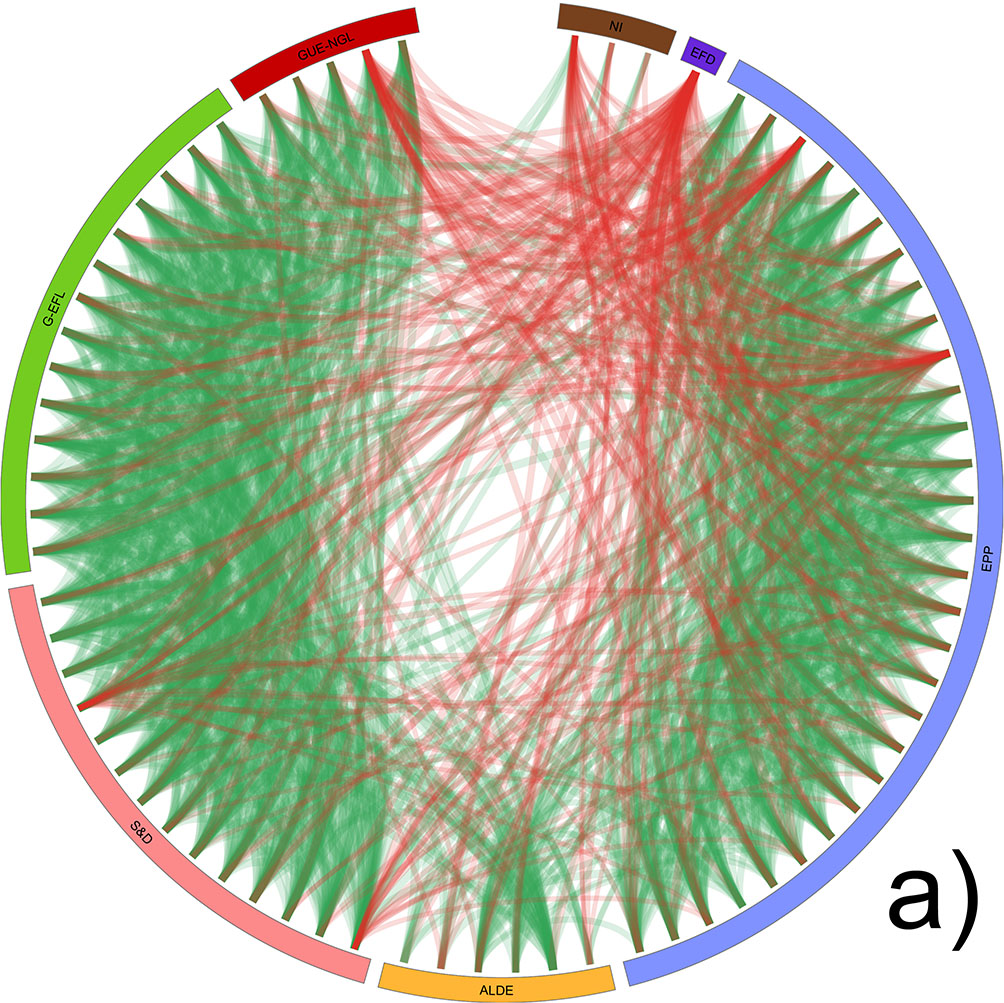}
    \includegraphics[width=0.25\textwidth]{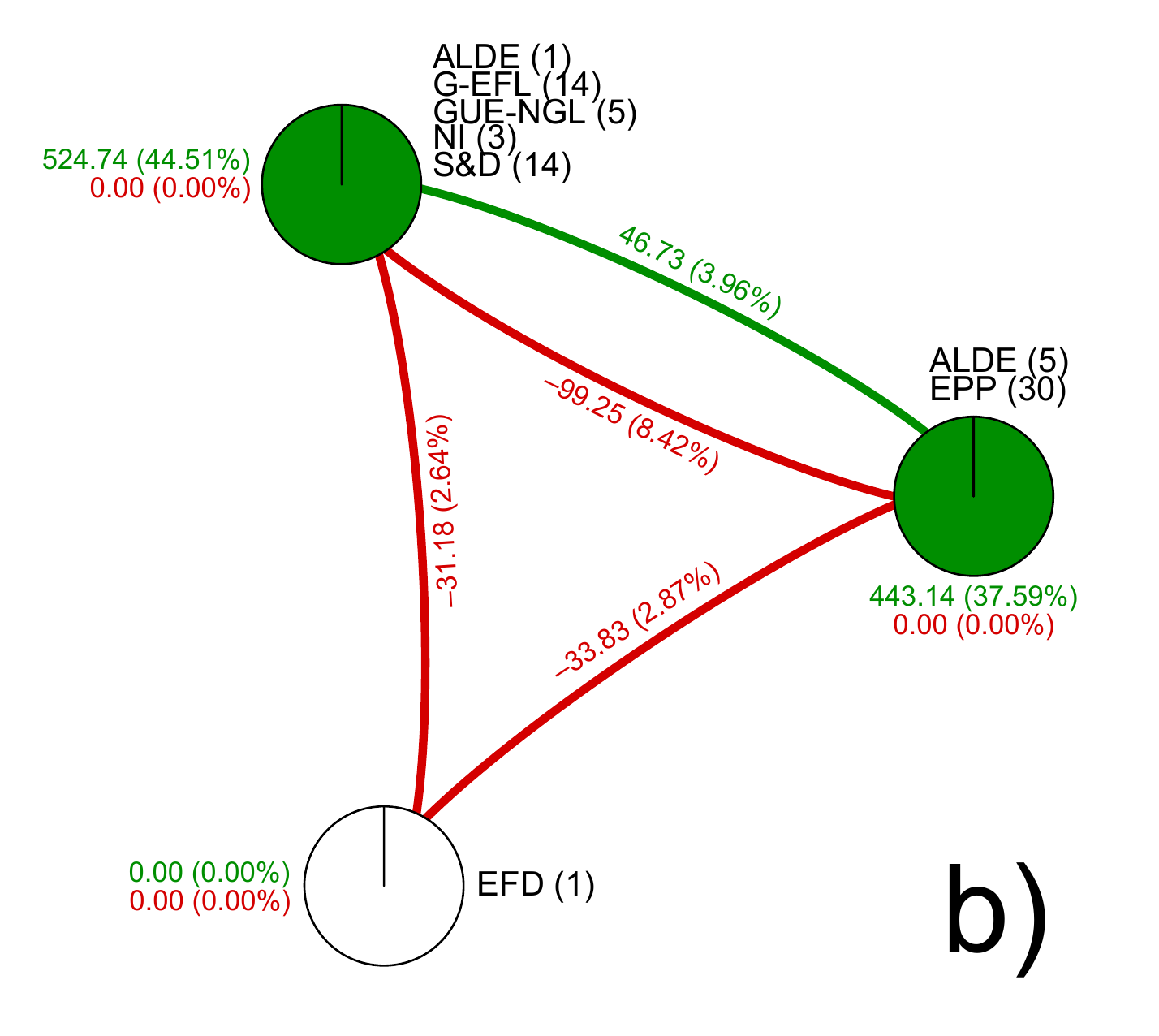}
    \includegraphics[width=0.25\textwidth]{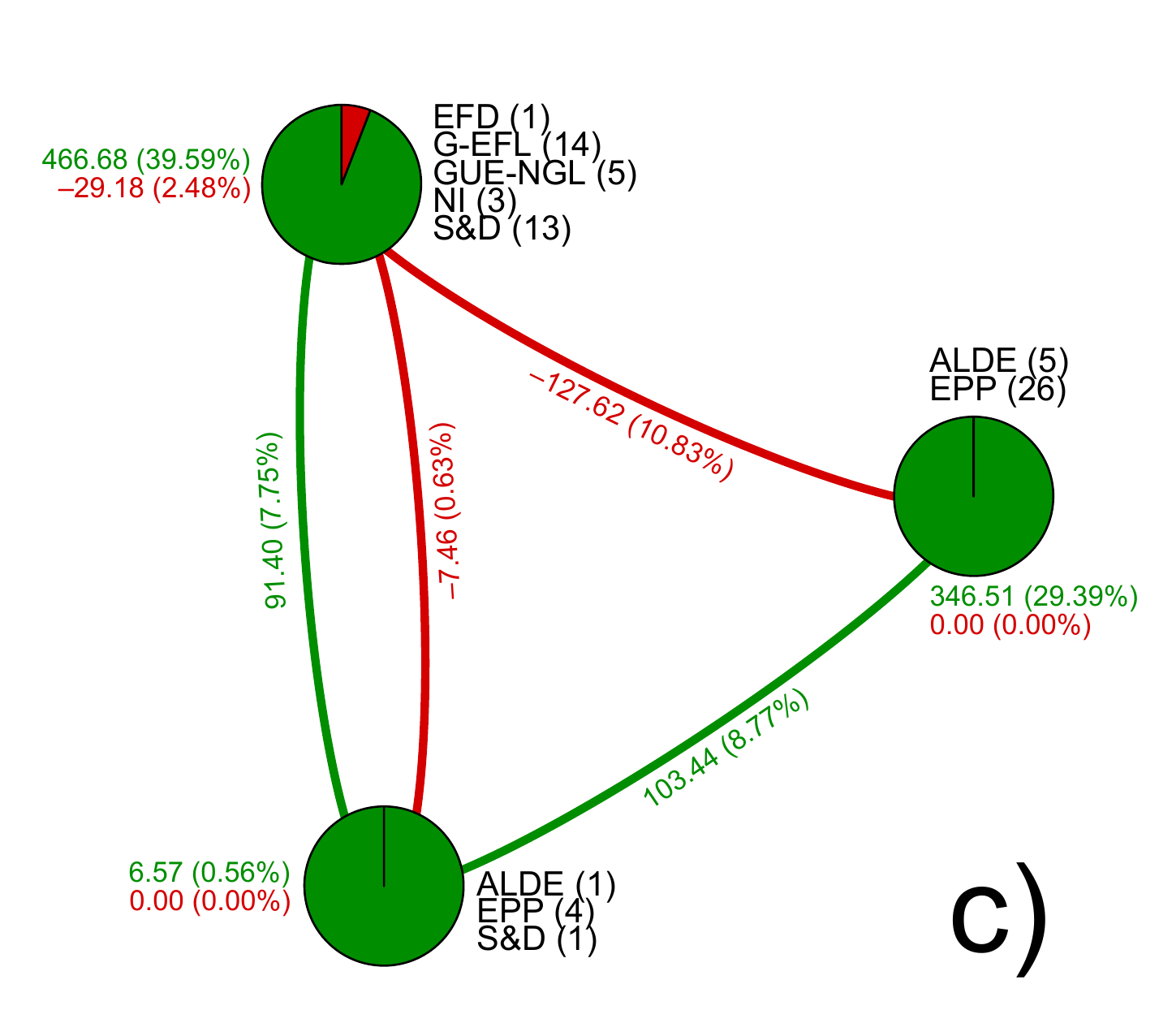}
    \includegraphics[width=0.25\textwidth]{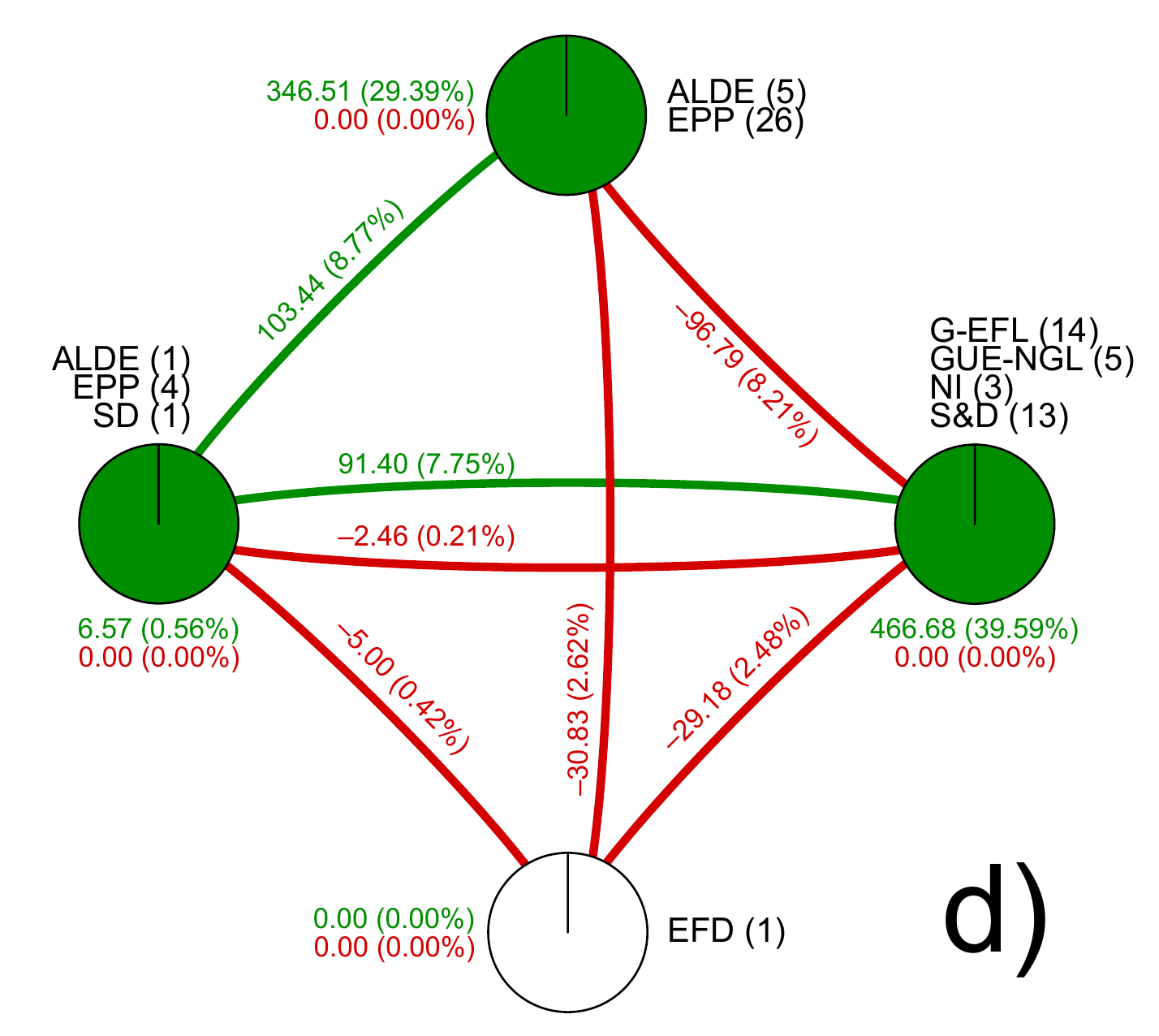}\\
    \includegraphics[width=0.21\textwidth]{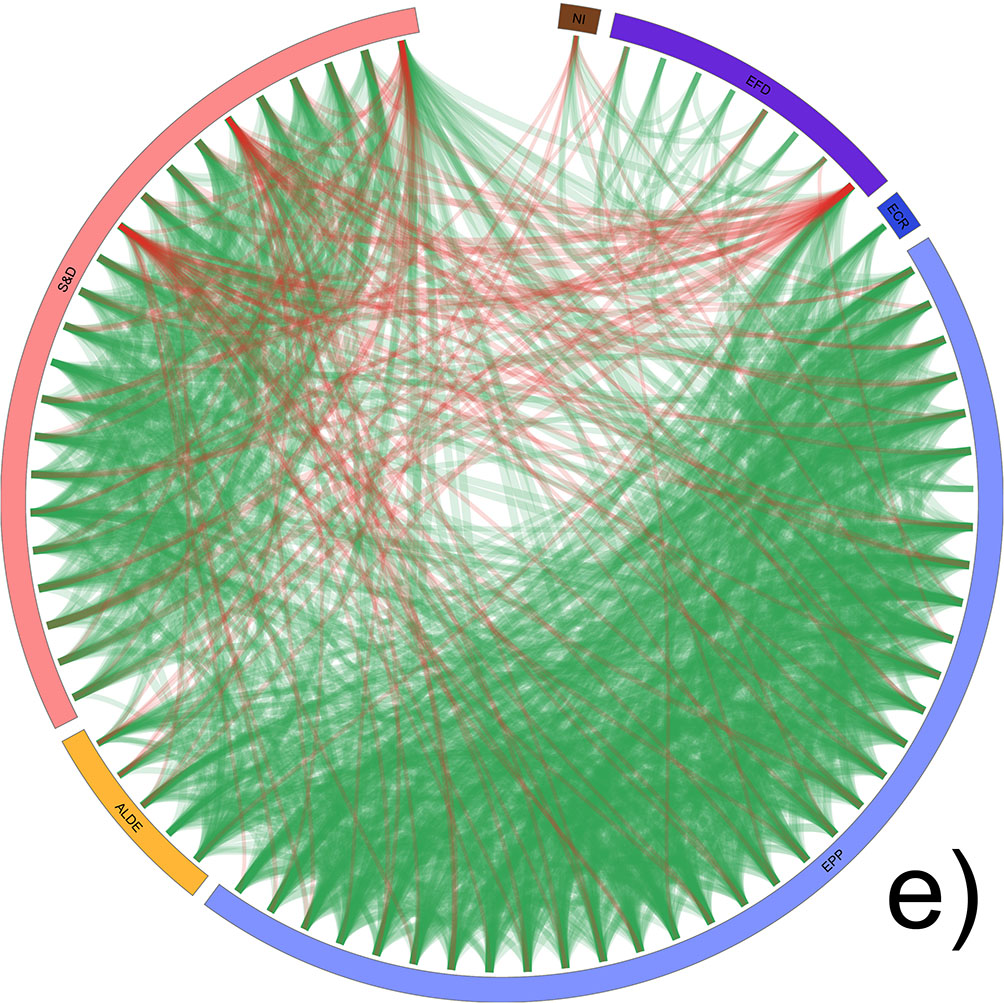}
    \includegraphics[width=0.25\textwidth]{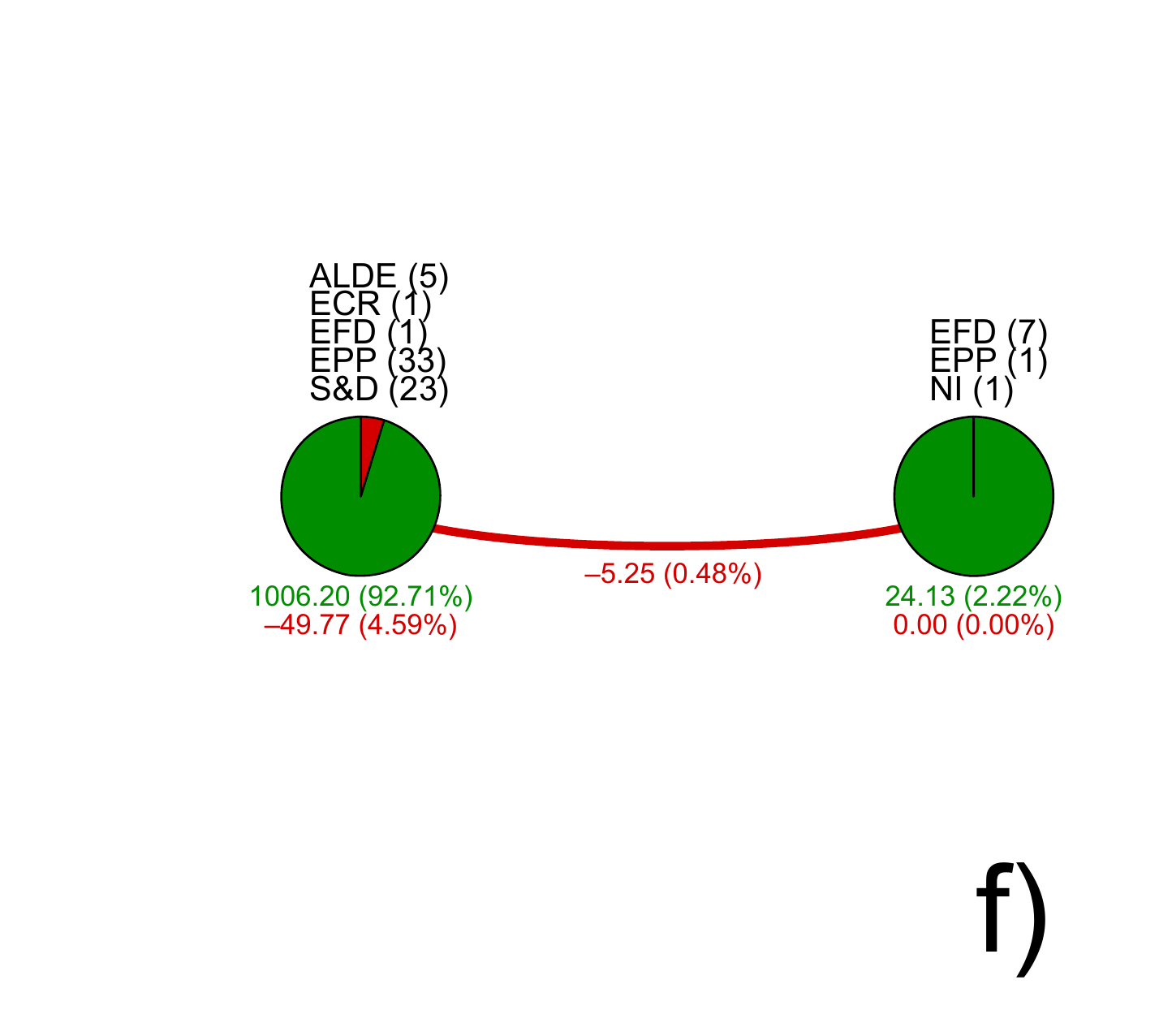}
    \includegraphics[width=0.25\textwidth]{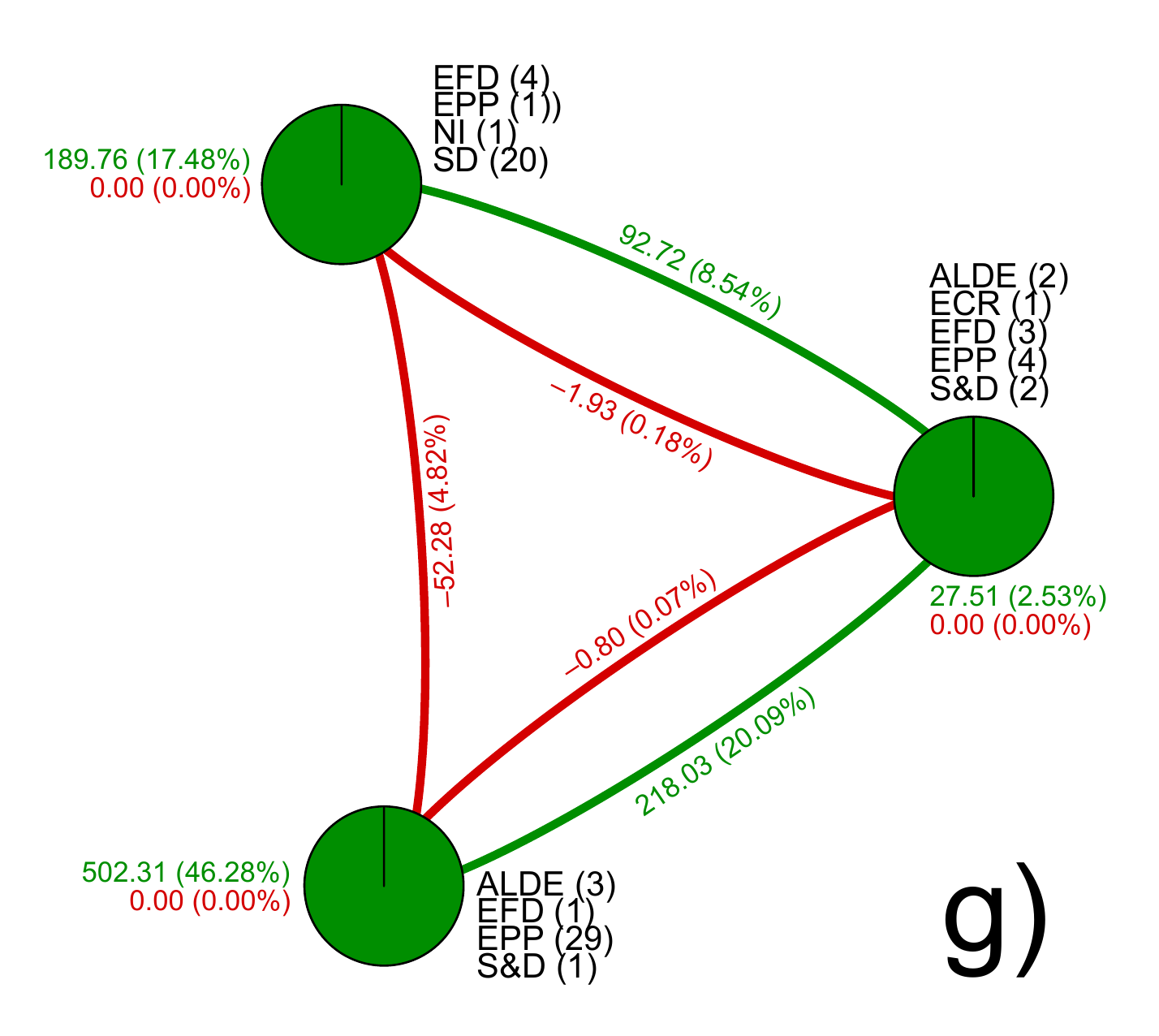}
    \includegraphics[width=0.25\textwidth]{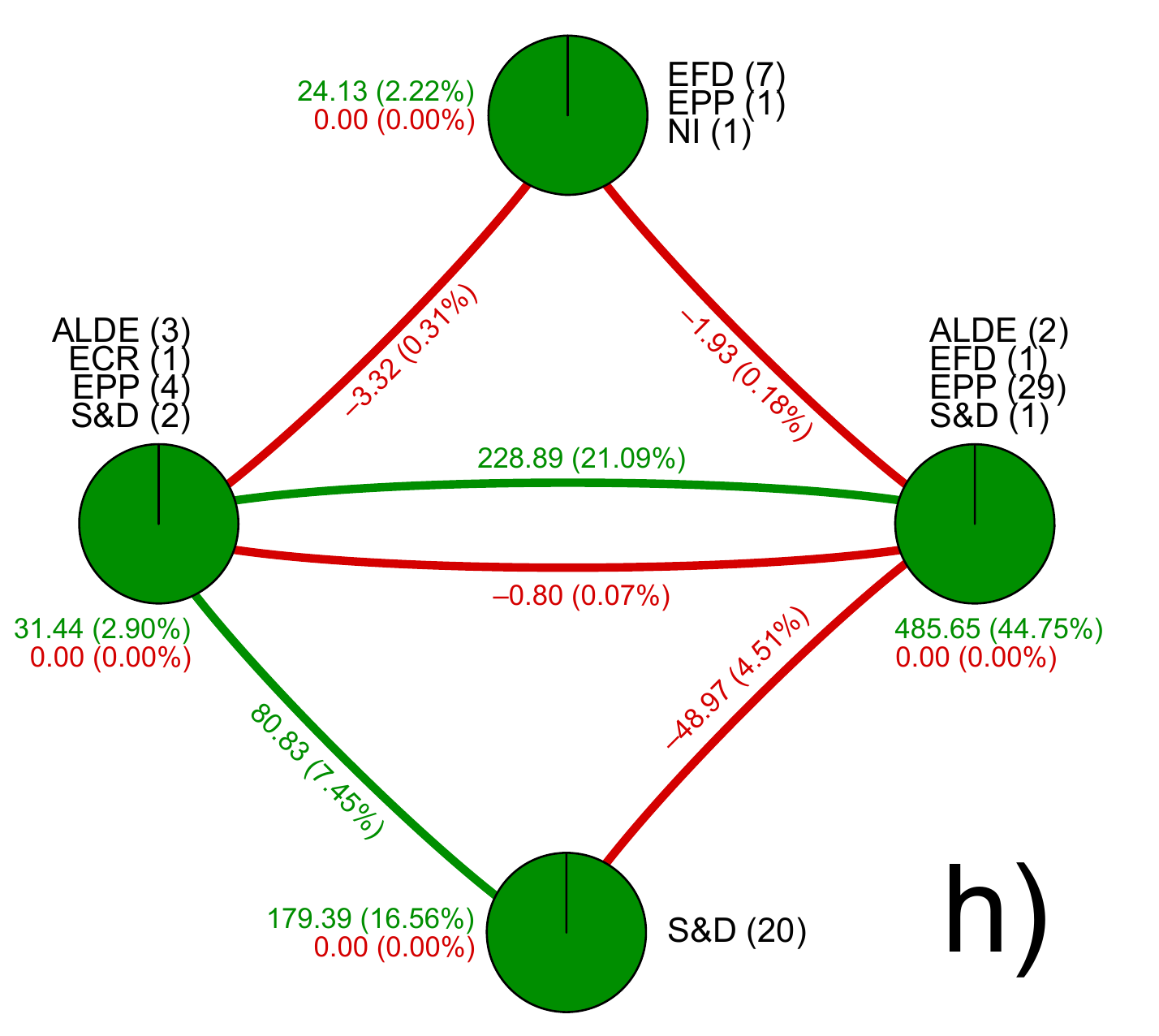}
	\caption{French (top) and Italian (bottom) MEPs for the ECON domain during the year 2009-10: individual networks (a) ; and cluster networks for the CC problem (b,f), the RCC problem with $k=3$ (c,g) and $k=4$ (d,h).}
	\label{fig:FrEcon}
\end{figure*}

We now turn to the Italian MEPs, which are represented by the bottom plots of Figure~\ref{fig:FrAgri}. first point to notice in the individual network (e) is the complete absence of any G-EFL or GUE-NGL MEPs, which seems to result in much fewer negative links, and therefore less polarization. The CC solution, represented in plot (f), contains $2$ clusters, with an imbalance of $8.58$. They are connected as much positively as negatively, and the smaller is a singleton corresponding to \textit{Gianni Vattimo}. All of his links are negative, and represent most of the negative links in the network. According to his biography\footnote{\url{https://en.wikipedia.org/wiki/Gianni_Vattimo}}, he has a very specific background and ideological position (nihilist philosopher, communist, speaks his mind) we can confidently state he is an outlier. If we ignore him, we can conclude that no polarization was detected by CC, probably due to the absence of environmentalists. 

Cluster graph (g) represents the partition obtained for RCC and $k=2$. The imbalance is slightly improved, and the partition is quite similar to that of CC, except Vattimo is joined by a few members of other groups. Interestingly, the resulting cluster has only negative internal links, whereas it is mostly positively connected to the other cluster. This is called \textit{internal hostility} by Doreian \cite{Doreian2009}. In this case, we suppose that this cluster gathers the most opposed MEPs from the main groups, and that they are likely to be agents of polarization, pulling their groups in opposite directions. 

The partition obtained for $k=3$ (d) has an almost-zero imbalance ($0.49$), and differs in two points with the previous one: 1) the large cluster is split in two: S\&D vs. right-wing groups ; and 2) Vattimo's cluster changes slightly, and as a result becomes internally positively connected. The cluster graph is quite similar in structure to the one obtained for France with RCC $k=3$, except this time $S\&D$ has positive connections with a right-wing cluster and a small heterogeneous cluster, themselves in opposition (instead of right vs. left/environmentalists). Overall, the RCC $k=3$ partition seems to be more informative than the RCC $k=2$, but the very distinct voting behavior of Vattimo might hide the simple fact that the Italian MEPs are not polarized on agricultural questions.

%%%%%%%%%%%%%%%%%%%%%%%%%%%%%%%%%%%%%%%%%%%%%%%%%%%%%%%%%%%%%%%
\subsection{Economic \& Monetary Affairs}
\label{sec:ResultsEcon}
The top plot in Figure~\ref{fig:FrEcon} display the ECON results for the French MEP during the year 2009-10. The individual network (a) contains much more negative links for AGRI. As we will see, this is meanly due to the fact that, this time, the two main antagonistic groups have sensibly the same size. The CC solution (b) contains $3$ clusters for a relatively high imbalance ($46.72$). There are $2$ large clusters: one dominated by left-wing groups (GUE-NGL, G-EFL, S\&D) and the other by right-wing ones (ALDE and EPP). Like for AGRI, we have some exceptions: NI (far right) and Lepage (ALDE) are placed in the left-wing cluster. For NI, we observe only a few positive links with the left, but many negative ones with the right, which is why they were put in the left-wing cluster. The third cluster is a singleton: de Villiers, like for AGRI, because he is only negatively connected to the other groups. The clusters constitute a purely negative triangle: all clusters are in strong opposition while internally positive. However, de Villiers is likely to be an outlier again, so this rather reflects a strong bipartition. This partition is consistent with the groups' ideologies regarding economics, with a clear left/right divide. It highlights the main difference with AGRI: this time, S\&D is on the side of G-EFL/S\&D (positive links) against EPP/ALDE (negative links), and does not constitute an intermediary cluster. 

The result obtained for RCC $k=3$ (c) has a slightly improved imbalance ($36.64$). The partition also exhibits a strong left/right divide, but not as strong as with CC. The third cluster does not contain de Villiers, and is instead a heterogeneous gathering of MEPs from $3$ distinct groups (EPP, S\&D and ALDE), which seem to be outliers in their own groups. For instance, it contains Lepage, which tends to vote like G-EFL when other ALDE MEPs votes like EPP. The cluster graph shows an imbalanced triangle, in which the outliers group is positively connected to both wings, whereas the other two clusters are very strongly negatively connected. Therefore, we consider this group has an intermediate position on the political spectrum, as previously observed for AGRI (with a different distribution, though).

When considering $k=4$ (d), the partition is the same except that de Villiers is placed in his own cluster. This small change causes the imbalance to drop to $2.5$. However, it is worth noticing that de Villiers was absent to many votes during this period ($27$ out of $35$), which explains the number and strength of his negative links. Due to these absences, these links are not very meaningful. So, in terms of interpretation, this partition is the same as the previous one, as shown by the cluster graph. Again, as observed for AGRI, a lower imbalance does not necessarily means the partition is more informative: it is necessary to go back to the original data to correctly interpret the results. 

We now turn to Italian MEPs, represented in the bottom plots of Figure~\ref{fig:FrEcon}. The individual network (e) contains much fewer negative links compared to the French network, and they are mostly attached to a few MEPs (especially \textit{Magdi Cristiano Allam} and \textit{Pino Arlacchi}). Like before, this is due to the fact these MEPs were absent to a lot of voting sessions, and always vote \textsc{Against}: by comparison, frequent voters tend to have their similarity scores smoothed when averaging over time. The CC solution (f) splits the network in $2$ clusters: one contains almost all far-right MEPs (EFD and NI), and the other the rest of the network. We clearly have $2$ antagonistic clusters, however this is due, in part, to the possibly overestimated negative links mentioned earlier. Overall, the network does not seem to display as much polarization as the French one.

The result obtained with RCC for $k=2$ is \textit{exactly} similar to that of CC, and for this reason it is not shown. For $k=3$ (g), the imbalance is almost zero ($0.49$), and the partition is very different: the large cluster is roughly split into a left-wing (S\&D) and a right-wing (EPP and ALDE) clusters, whereas the third group is very heterogeneous, and contains MEPs from all groups except NI. The cluster graph takes the form of the triangle already observed before: two opposed clusters (here: left vs. right) both positively connected to the third one (here: the heterogeneous cluster), which holds an intermediary position. With $k=4$ (h), we get a very similar partition, with an extra cluster gathering the same frequently absent far-right MEPs. As mentioned before, this group is likely to be an artifact due to the way absence is handled during the extraction process. However, RCC must be credited for being able to identify this cluster of interest while still distinguishing the three other political trends present among the Italian MEPs.

\todoVL{ideas to put somewhere: the idea the optimal solution is not necessarily the most relevant is not natural in operations research. Maybe also say that the problem could need to be reformulated to fit the application goals? also, do we speak about the multiplicity of optimal solutions?}

\todoVL{there is a sensitivity to absences in the extraction process. RCC kind of makes up for it, but it's be more efficient to correct it in the extraction phase.}

\todoVL{the fact we get triangular cluster graphs is due to the fact the networks get averaged over several texts (otherwise, only 2 clusters if consider a single text). also, these triangles correspond to a situation where conflict is diffused in the graph, otherwise the bipolarity would persist over all texts. }

\todoVL{we often see a new cluster appearing when going from CC to RCC. it can either be heterogeneous , in which case it is either an artefact (absence) (or ...?), or be homogeneous, in which case it is an intermediary cluster, on the political spectrum. put that with the triangle story about the average and so on.}

\todoVL{Should we insist more on the complementary nature of CC and RCC ? (no additional computation, since RCC requires CC)}

%%%%%%%%%%%%%%%%%%%%%%%%%%%%%%%%%%%%%%%%%%%%%%%%%%%%%%%%%%%%%%%
%%%%%%%%%%%%%%%%%%%%%%%%%%%%%%%%%%%%%%%%%%%%%%%%%%%%%%%%%%%%%%%
\section{Conclusion}
\label{sec:Conclusion}
In this article, we extract and analyze signed networks representing the voting behavior of MEPs, using two graph partitioning methods for the CC and RCC problems. We first propose a filtering step to make the original networks sparser, which eases both their processing and the interpretation of the results. We show empirically that the effect of this step on the obtained partitions is negligible. We also assess the quality of a heuristic allowing to speed up the partitioning process, and show it is significantly faster than an exact method, while identifying solutions of the same quality. Finally, we focus on $4$ networks of interest from our dataset, and validate the relevance of the partitioning algorithms by performing a qualitative evaluation of their results. We show that the two partitioning approaches are complementary, and allow to identify various types of clusters: some match the traditional ideological divides between the political groups, whereas other highlights specific behaviors or positions in the EP. 

However, we also identified some limitations, which we plan to solve in our future work. First, some artifacts appear when certain MEPs combine two specific behaviors: frequent absences and systematic \textsc{Against} vote. The partitioning tools are able to detect these cases, but they still obfuscate the results, so this issue has to be solved before partitioning. Second, our interpretation was limited to our short knowledge in the domain of European politics: we need to collaborate with a specialist, in order to further assess the quality of the partitioning methods.

%%%%%%%%%%%%%%%%%%%%%%%%%%%%%%%%%%%%%%%%%%%%%%%%%%%%%%%%%%%%%%%
%%%%%%%%%%%%%%%%%%%%%%%%%%%%%%%%%%%%%%%%%%%%%%%%%%%%%%%%%%%%%%%

\begin{acks}
  This research benefited from the support of the Agorantic FR 3621, as well as the \textit{FMJH Program Gaspard Monge in optimization and operation research} (Project 2015-2842H), and from the support from EDF to this program.
\end{acks}

%%%%%%%%%%%%%%%%%%%%%%%%%%%%%%%%%%%%%%%%%%%%%%%%%%%%%%%%%%%%%%%
%%%%%%%%%%%%%%%%%%%%%%%%%%%%%%%%%%%%%%%%%%%%%%%%%%%%%%%%%%%%%%%
\bibliographystyle{ACM-Reference-Format}
\bibliography{biblio} 

\end{document}